\documentstyle[aps,prl,tighten,eqsecnum]{revtex}

\input epsf

\def\opsimeq{\mathop{\simeq}}

\def\opsim{\mathop{\sim}} 

\begin{document} 
\draft 
\preprint{}

\title{\bf  ON THE LOCALIZATION OF RANDOM HETEROPOLYMERS
 AT THE INTERFACE BETWEEN TWO SELECTIVE SOLVENTS}

\author{C\'ecile Monthus}

\address{ CNRS-Laboratoire de Physique Th\'eorique et Mod\`eles Statistiques \\
DPT-IPN, B\^at. 100, 91400 Orsay, France, FRANCE \\ e-mail : monthus@ipno.in2p3.fr } 

\date{\today}

 \maketitle 
\begin{abstract}
To study the localization of random heteropolymers at
 an interface separating two selective solvents within 
the model of Garel, Huse, Leibler and Orland,
Europhys. Lett. {\bf 8} 9 (1989), we propose a
 disorder-dependent real space renormalization approach.
 This approach allows to recover
 that a chain with a symmetric distribution in hydrophobic/hydrophilic components is localized at any temperature in the thermodynamic limit,
whereas a dissymmetric distribution in hydrophobic/hydrophilic components
leads to a delocalization phase transition. It yields
in addition explicit expressions for thermodynamic
quantities as well as a very detailed description of the statistical
properties of the heteropolymer conformations
 in the high temperature limit. In particular, 
scaling distributions are given for the lengths of the blobs
in each solvent, for the polymer density, and for some
correlation functions.
In the case of a small dissymmetry in hydrophobic/hydrophilic components,
 the renormalization approach yields explicit expressions 
for the delocalization transition 
temperature and for the critical behaviors of various quantities :
in particular, the free energy presents an essential singularity at the transition (the transition is thus of infinite order),
 the typical length of blobs in the preferred
solvent diverges with an essential singularity, whereas
the typical length of blobs in the other
solvent diverges algebraically.
 Finite-size properties are also characterized
in details in both cases. In particular, 
we give the probability distribution
of the delocalization temperature for the ensemble
of random chains of finite (large) length $L$, and the distribution
of the numbers of blobs for the chains that are still localized at
a given temperature. 
Finally, we discuss the non-equilibrium dynamics at temperature $T$ 
starting from a zero-temperature initial condition.
\end{abstract}

\section{Introduction}

Among the various subjects concerning the physics of polymers,
the behavior of heteropolymers containing hydrophobic
and hydrophilic components in solvents are of particular
interest since they have obvious importance in biology \cite{review1}.
It is well known for instance that in a polar solvent, these heteropolymers
prefer conformations where the hydrophilic components
are in contact with the polar solvent, whereas hydrophobic
components avoid contacts with the solvent.
The behavior of heteropolymers
in the presence of an interface 
separating two selective solvents, one favorable to the hydrophobic
components and the other to the hydrophilic components,
is less obvious, and has been much studied recently. 
In the pioneering work of Garel, Huse, Leibler and Orland \cite{garel},
a model was proposed and studied via Imry-Ma arguments, 
an analysis of the replica Hamiltonian and numerics : it was found
that a chain with a symmetric distribution in hydrophobic/hydrophilic
 components is always localized around the interface
at any temperature (in the thermodynamic limit), whereas a chain
 with a dissymmetric distribution in hydrophobic/hydrophilic
 components presents a phase transition separating
a localized phase at low temperatures from a delocalized phase
into the most favorable solvent at high temperatures.
 Experimentally, the presence of
copolymers was found to stabilize the interface
between the two immiscible solvents \cite{experiments},
since the localization of the heteropolymers at the interface
reduces the surface tension.
By now, the predictions of Reference \cite{garel} have been confirmed 
in the physics community by various approaches 
including molecular dynamics simulations \cite{yeung},
Monte Carlo studies \cite{sommer},
variational methods for the replica Hamiltonian
\cite{stepanow} \cite{maritanreplica}, and exact bounds for the free-energy \cite{maritanbounds}. Mathematicians have also been interested
in this model, but the exact results obtained up to now are still far
from a complete explicit solution.
The localization at all temperatures for the symmetric case
was proven in References \cite{sinai} \cite{albeverio} by various criteria
of localization on the paths measure. In the dissymmetric case,
the existence of a transition line in temperature vs dissymmetry plane
was proven in Reference \cite{bolthausen} where
some bounds for the free-energy were obtained in the asymptotic
regimes of low and high temperatures. Finally,
the relations between various localization criteria
concerning either free-energy or path properties have been studied
in Reference \cite{biskup}.

In this paper, we propose a new approach to study the model
of Garel et al. \cite{garel}, based on a disorder-dependent
real space renormalization procedure. Since this type of approach
is not usual in the context of 
the thermodynamics of classical disordered systems, 
it seems useful to recall how these methods
 have been successfully applied in other domains. 
Disordered-dependent real space renormalization 
approaches have first been introduced 
in the field of disordered quantum spin chains \cite{Ma}
\cite{fisher} \cite{fisheryoung} \cite{hyman} \cite{spin1}.
More recently they have been used to study random walkers
in 1D disordered environments \cite{RGletter} \cite{RGSinai}, 
reaction-diffusion in 1D disordered environments \cite{RGreadiff} and
non-equilibrium dynamics of disordered classical spin chains \cite{RGletter}
\cite{RGRFIM}. A similar method has also been introduced
independently to study the coarsening of the pure one-dimensional $\Phi^4$
model at zero temperature \cite{phi4}.
In all these fields, these methods, which are not
 ``exact from first principles",
 have however been remarkably successful in reproducing the exact results
that were already known (see References \cite{fisher} \cite{RGSinai}),
and in producing a lot of novel exact results
for exponents and scaling functions for a large
variety of physical quantities. These results apply
 in the large renormalization
scale regime, corresponding respectively to low-temperature
in the field of disordered quantum spin chains, and to large time
behaviors in the other fields concerning dynamics.
These new predictions have moreover been
confirmed numerically whenever they have been tested 
\cite{fisheryoung} \cite{testRGnume} \cite{chave}.

For the random heteropolymer problem, the disorder-dependent
real space renormalization procedure that we propose
in this paper is defined 
 to select the ``important" configurations for the thermodynamics
of a chain with a given realization of the 
disordered sequence in hydrophobic/hydrophilic components
at a given temperature. 
We expect this renormalization approach to become
accurate in the large renormalization
scale regime, corresponding here to the region of high temperatures.
Since the localized phase extends up to a critical temperature
$T_c$ that goes to infinity as the dissymmetry parameter
of the distribution in hydrophobic/hydrophilic components
goes to zero, our approach allows to study the localized phases
of symmetric chains and slightly dissymmetric chains.

The paper is organized as follows. 
In the remainder of the Introduction, we
recall the model of Garel et al. \cite{garel}
 in Section \ref{model}, and give a summary
of the main results of the present paper in Section \ref{summary}.
In Section \ref{physical}, we recall the physics
of the heteropolymer problem at $T=0$ and at high temperatures,
in order to motivate the introduction of
a disorder-dependent
real-space renormalization procedure
 in Section \ref{RG}.
The results of the renormalization approach
 are given for symmetric and dissymmetric
chains in Sections \ref{symmetric} and \ref{biased} respectively.
In Section \ref{finite}, we characterize the finite-size properties 
of the problem by considering chains of finite (large) length.
In Section \ref{dynamics}, we discuss how the renormalization procedure
also describes the non-equilibrium dynamics of a random chain
at high temperature starting at $t=0$ from a zero-temperature 
initial condition. Finally Section \ref{conclusion} contains
the conclusions. In the Appendix \ref{appendixImryMa}, we compare
some results of the renormalization approach
with the corresponding results obtained by Garel and Orland
via an extension of the usual Imry-Ma argument \cite{garelImryMa}.

\subsection{Model and notations}

\label{model}

We consider the model introduced by Garel et al. \cite{garel} 
with some slightly different notations:
a polymer chain consists of $L$ monomers indexed by $i=1,2, \ldots, L$,
where each monomer $i$ carries a quenched random charge $q_i$.
The $\{q_i\}$ are independent identical random variables 
drawn with some probability distribution $C(q)$. 
An interface at $z=0$ separates 
a solvent in the domain $z > 0$ favorable to positive charges $q>0$,
from a solvent in the domain $z < 0$ favorable to negative charges $q<0$.
More precisely, the energy of a configuration ${\cal C}=\{{\vec r}_i=
(x_i,y_i,z_i)\}$
of the heteropolymer characterized by a realization $\{q_1,q_2, \ldots q_L\}$ reads
\begin{eqnarray}
E({\cal C})=-\sum_{i=1}^L q_i  {\rm sgn}(z_i) 
\qquad ,
\end{eqnarray}
and the partition function reads
\begin{eqnarray} \label{partition}
Z_L(\beta;\{q_i\}) = {\rm Trace}_{\{{\vec r}_i\}}
\prod_{i=1}^{L-1} \delta \left( \vert {\vec r}_{i+1}-{\vec r}_i \vert -a \right)
\exp \left( \beta \sum_{i=1}^L q_i  {\rm sgn}(z_i) \right)
\qquad .
\end{eqnarray}
In a continuum version of this model, the monomer index $i$ becomes
a continuous variable $s$, and the $(x,y)$ coordinates
play no role so that the partition function becomes a path
integral over one-dimensional Brownian trajectories $\{z(s)\}$
\cite{garel}
\begin{eqnarray} \label{functional}
Z_L(\beta;\{q(s)\}) = \int {\cal D} z(s)
\exp \left( - \frac{1}{2 D} \int_0^L ds \left( \frac{dz}{ds}\right)^2 
+ \beta \int_0^L ds q(s) {\rm sgn }(z(s)) \right)
\qquad .
\end{eqnarray}

In this paper, we propose a disorder-dependent real-space renormalization 
procedure, that should give valid results in the universal
regime of large scales, where the details of the microscopic
charge distribution $C(q)$ are not important, as long as
the distribution $C(q)$ satisfy the hypothesis of the Central
Limit Theorem. 
The important parameters of $C(q)$ will thus be the mean value
and the variance
\begin{eqnarray} \label{defq}
&& \overline{q}= \int_{-\infty}^{+\infty} dq \ q \ C(q)=q_0 \\
&& \overline{q^2}-q_0^2=\int_{-\infty}^{+\infty} dq \ (q-q_0)^2 \ C(q)
= 2\sigma
\qquad .
\end{eqnarray}
To study the biased case $q_0 \neq 0$, we will always choose
the convention $q_0 >0$ and introduce the parameter
 $\delta$ defined as the non-vanishing 
root of the equation \cite{RGSinai}
\begin{eqnarray} \label{defdelta}
\overline{e^{- 2 \delta q}} 
\equiv \int_{-\infty}^{+\infty} dq e^{- 2 \delta q} C(q)= 1
\qquad .
\end{eqnarray}
For instance, in the case of a Gaussian distribution 
considered in References \cite{garel} \cite{stepanow} \cite{maritanreplica} \cite{maritanbounds}
\begin{eqnarray} \label{gauss}
C(q)=\frac{1}{\sqrt{4 \pi \sigma}} e^{- \frac{(q-q_0)^2}{4 \sigma} }
\qquad ,
\end{eqnarray}
the parameter $\delta$ is simply the ratio between 
the mean value $q_0$ and the variance $(2 \sigma)$
\cite{fisher}
\begin{eqnarray}
 \delta=\frac{ q_0}{ 2 \sigma}  \qquad .
\end{eqnarray}

For the case of binary distribution considered in References \cite{sommer} \cite{maritanbounds} \cite{bolthausen} \cite{biskup} 
where the charge $q$ takes a positive
value $q_+$ with probability $c$ and
a negative value $(-q_-)$ with probability $(1-c)$ then
\begin{eqnarray} 
&&  q_0= c q_+ -(1-c) q_-  \qquad , \\
&& \sigma= \frac{c(1-c)}{2} (q_+ +q_-)^2 
\qquad ,
\end{eqnarray}
and $\delta$ is the solution of
\begin{eqnarray} \label{deltabinary}
c e^{- 2 \delta q_+} +(1-c) e^{+ 2 \delta q_-} =1
\qquad .
\end{eqnarray}
For instance in the case $q_-=q_+=q_1$, we have
\begin{eqnarray} \label{binarypm}
&& q_0=(2c-1)q_1 \\
&& \sigma=2 c (1-c) q_1^2 \\
&& \delta=\frac{1}{ 2 q_1} \ln \frac{c}{1-c}
\qquad .
\end{eqnarray}

An important property of the parameter $\delta$
is that it becomes universal and coincides with the Gaussian expression 
 in the limit of small dissymmetry $\delta \to 0$ \cite{fisher}
\begin{eqnarray} \label{univ}
\delta \simeq \frac{q_0}{2 \sigma}
\qquad .
\end{eqnarray}

Concerning the spatial behavior of the polymer, 
the only important parameter will be
the diffusion coefficient $D$ characterizing the large distance behavior
of the free chain 
\begin{eqnarray}
< (z(s)-z(s'))^2 > \simeq D \vert s-s' \vert
\qquad .
\end{eqnarray}
We stress that in this paper, we do not take into account
excluded volume effects because they do not have any crucial
effect for the problem of the localization of the heteropolymer
at the interface \cite{garel} \cite{sommer} \cite{maritanbounds}.

\subsection{Quantities of interest and summary of main results}

\label{summary}

The quantities of interest 
to characterize the localized phase of heteropolymers around the interface
are on one hand the free-energy $f(T)$ per monomer that governs
the thermodynamic, and on the other hand 
the statistical properties of the spatial conformations of the chain.
In particular, the important characteristic lengths 
are the typical lengths $l^{\pm}_{ blob}(T)$ of blobs in the $(\pm)$ solvents 
 as well as the typical distances $z^{\pm}_{ blob}(T)$ between the chain
and the interface. At a more refined level, 
one may also consider the probability
distributions $P^{\pm}(l)$ of the lengths of $(\pm)$ blobs,
the density $\rho^{\pm}(z)$
of polymer at a distance $z$ from the interface in the $(\pm)$ solvents,
and even correlation functions as for instance
the solvent-solvent correlation
$\overline{< {\rm sgn} (z(s)) {\rm sgn} (z(s')) > }$.
Finally, we will also be interested in the finite size effects
of the problem. 

Before introducing our approach and proceeding with
the calculations of these quantities of interest, we now summarize  
the main results that will be derived from the real-space
renormalization approach in the remainder of the paper.

\subsubsection{Summary of main results for the symmetric case ($q_0=0$)}

For the case of symmetric chains ($q_0=0$), 
the renormalization approach yields the 
following behaviors at high temperature
\begin{eqnarray}
&& f(T) \sim - \frac{ \sigma}{  T \ln T} \qquad , \\
&& l_{ blob}(T) \sim \frac{T^2}{\sigma} (\ln T)^2 \qquad ,\\
&& z(T) \sim \frac{T}{\sqrt{\sigma} } (\ln T) \qquad ,
\end{eqnarray}
in agreement with the Imry-Ma argument results \cite{garel}
(see also Section \ref{highT} for the discussion of the presence
of the logarithmic correction).
In addition, we obtain that the rescaled length of blobs
\begin{eqnarray}
 \lambda=\frac{ \sigma l}{ 9 T^2 (\ln T)^2}
\end{eqnarray}
is distributed with the law 
\begin{eqnarray}  
  P(\lambda)  
&& = \sum_{n = -\infty}^{\infty} \left(n+\frac{1}{2}\right)
\pi (-1)^n e^{- \pi^2 \lambda \left(n+\frac{1}{2}\right)^2} 
\opsimeq_{\lambda \to \infty} 
\pi  e^{- \frac{\pi^2}{4} \lambda } \\
&& = \frac{1}{\sqrt \pi \lambda^{3/2}}
\sum_{m = -\infty}^{\infty} (-1)^m (m+\frac{1}{2})
 e^{-  \frac{1}{\lambda} (m+\frac{1}{2})^2}
 \opsimeq_{\lambda \to 0} 
 \frac{1}{\sqrt \pi \lambda^{3/2}}
 e^{-  \frac{1}{ 4 \lambda} }
\qquad ,
\end{eqnarray}
The polymer density takes the scaling form $\rho(z) dz = R(Z) dZ $ where
the rescaled distance to the interface
\begin{eqnarray}
Z=  \sqrt{\frac{2 \sigma }{D}} \frac{z}{ 3 T (\ln T)}
\end{eqnarray}
is distributed with the following 
law $R(Z)$ that decays exponentially at large distance
\begin{eqnarray}
 R(Z)= 4 \int_0^{\infty} d \lambda P(\lambda) \sqrt{\lambda}
\int_{\frac{Z}{\sqrt{\lambda}}}^{\infty} du e^{- u^2}
 \opsimeq_{Z \to \infty}  \frac{8}{\pi} \sqrt{2 Z} e^{-\pi Z} \qquad .
\end{eqnarray}
Finally, to characterize the finite size effects,
we will compute the distribution of
 the delocalization temperature $T_{\rm deloc}$ 
over the ensemble of random cyclic finite chains of (large) length $L$.
The result is that the rescaled variable 
\begin{eqnarray}
g=\frac{ 3 }  {\sqrt{\sigma L}} T_{\rm deloc} \ln T_{\rm deloc}
\end{eqnarray}
 is distributed with the law
\begin{eqnarray}  
D(g)   
&& = \frac{\pi^2}{g^3}\sum_{n=1}^{+\infty} (-1)^{n+1} 
n^2 e^{-   \frac{n^2 \pi^2}{4 g^2}} 
\opsimeq_{g \to 0} 
\frac{\pi^2}{g^3}  e^{-   \frac{ \pi^2}{4 g^2}}\\
&& = \frac{2}{\sqrt{\pi }} 
\sum_{m=-\infty}^{+\infty} 
\left[ 2 (2m+1)^2 g^2 -1 \right] e^{-  (2m+1)^2 g^2}
\opsimeq_{g \to \infty} 
\frac{4}{\sqrt{\pi }} g^2  e^{-   g^2}
\qquad .
\end{eqnarray}

\subsubsection{Summary of main results for the dissymmetric case ($q_0>0$)}

For the case of dissymmetric chains ($q_0 >0$), 
in the limit $\sigma \gg q_0$,
the renormalization approach yields that the delocalization transition takes place at the critical temperature 
\begin{eqnarray}  
T_c=\frac{4 \sigma }{ 3 q_0} \qquad ,
\end{eqnarray}
in agreement with the scaling obtained previously by other methods
\cite{garel} \cite{stepanow} \cite{maritanreplica} \cite{maritanbounds}.
In addition, we obtain that the transition is of infinite order,
with the following essential singularity for the free energy
the free-energy $f(T)$  
\begin{eqnarray}  
f(T) -f(T_c) \opsimeq_{T \to T_c^-} 
- 2 q_0 \left(\ln \frac{4 \sigma}{q_0}  \right) 
\exp \left[ - \frac{ \ln \frac{4 \sigma}{q_0} }
{ \left(1-\frac{T}{T_c} \right)} \right]
\qquad .
\end{eqnarray}
For the statistical properties of the heteropolymer chain,
we find that the typical length $l^{+}_{ blob}(T)$ of blobs
in the preferred solvent diverges with an essential singularity
at the transition, whereas the typical length $l^{-}_{ blob}(T)$
in the other solvent diverges algebraically
\begin{eqnarray}
&& l^{+}_{ blob}(T) \opsimeq_{T \to T_c^-} 
\frac{\sigma}{q_0^2}  \exp \left[ + \frac{ \ln \frac{4 \sigma}{q_0} }
{ \left( 1-\frac{T}{T_c} \right)} \right] \\
&& l^{-}_{blob}(T) \opsimeq_{T \to T_c^-} 
\frac{\sigma}{q_0^2} \frac{ \ln \frac{4 \sigma}{q_0} }
{\left(1-\frac{T}{T_c} \right)} 
\end{eqnarray}
The rescaled length of blobs in the preferred solvent, defined as
\begin{eqnarray}
\lambda_+=\frac{l_+}
{ 
\frac{\sigma}{q_0^2} 
 \exp \left[
  \frac{ \ln \frac{4 \sigma}{q_0} }  
{ \left( 1-\frac{T}{T_c} \right)} 
\right] } 
\end{eqnarray}
is distributed with the exponential law $e^{-\lambda_+}$ near the transition.
The polymer density $\rho^+(z)$ takes the scaling form 
$\rho^+(z) dz = R^+(Z) dZ $ where the rescaled distance to the interface
\begin{eqnarray}
 Z=  \frac{  z}{ \sqrt{ \frac{ D}{2} } \frac{\sigma}{q_0} 
\exp \left[ + \frac{ \ln \frac{4 \sigma}{q_0} }
{ 2 \left( 1-\frac{T}{T_c} \right)} \right] }
\qquad 
\end{eqnarray}
is distributed with the following scaling distribution
that decays exponentially at large distance
\begin{eqnarray}
R^+(Z)= 4 \int_0^{\infty} du e^{- u^2}
\int_{\frac{Z}{u}}^{\infty} dv v^2 e^{- v^2} \opsimeq_{Z \to \infty}
 \sqrt{\pi Z} e^{-2 Z}
\qquad .
\end{eqnarray}
Finally, to characterize the finite size effects,
we will compute the distribution of
 the delocalization temperature $T_{\rm deloc}$ 
over the ensemble of random cyclic finite chains of (large) length $L$.
The result is that the random variable 
\begin{eqnarray}  
r=\frac{\sigma}{q_0^2 L} 
\left( \frac{4 \sigma}{q_0^2 } \right)^{\frac{T_{\rm deloc}}{T_c-T_{\rm deloc}}}
\end{eqnarray}
 is distributed with the law
\begin{eqnarray}  
D^+(r)= \frac{1}{r^2} e^{-\frac{1}{r}} 
\qquad .
\end{eqnarray}
In particular, the typical value for the delocalization temperature
presents a correction of order $(1/\ln L)$ with respect
the critical temperature $T_c$
\begin{eqnarray}  
 T_{\rm deloc}^{typ} \sim T_c \left(1-\frac{4 \sigma }{q_0^2 \ln L} \right)
\qquad .
\end{eqnarray}

\section{ Physical picture at $T=0$ and at high temperatures}

\label{physical}

In this section, we recall the physics of
the heteropolymer problem at $T=0$ and 
the Imry-Ma picture of Reference \cite{garel} valid at high temperature.
They will be useful to motivate the
introduction of the real-space renormalization procedure
in the next section.

\subsection{ Description of Ground States}

\label{ground}

At $T=0$, for a given realization $\{q_1,q_2, \ldots q_L\}$, 
each monomer $i$ will be in its preferred solvent according to
\begin{eqnarray}
{\rm sgn}(z_i)={\rm sgn}(q_i) \qquad .
\end{eqnarray}
The ground states have thus for energy
\begin{eqnarray}
E_0=-\sum_{i=1}^L \vert q_i \vert \qquad .
\end{eqnarray}
The corresponding configurations of the chain are all the random walks
that cross the interface each time
there is a change of signs in the realization $\{q_1,q_2, \ldots q_L\}$.
More precisely, it is convenient to
 decompose the realization $\{q_1,q_2, \ldots q_L\}$
into groups $(\alpha)$ of consecutive charges of the same sign,
each group $(\alpha)$ containing
a number $l_{\alpha}=1,2, \ldots$ of monomers,
and carrying an absolute charge $Q_{\alpha}$ with
\begin{eqnarray}
&& l_{\alpha} = \sum_{i \in \alpha} 1  \\
&& Q_{\alpha} = \vert \sum_{i \in \alpha} q_i \vert
\qquad .
\end{eqnarray}
Then the ground states of the chain can be decomposed
 into blobs $\alpha$ containing $l_{\alpha} \sim O(1)$ monomers.
The chain is thus localized around the interface
with a typical distance of order $O(1)$.

\subsection{ Imry-Ma arguments at high temperature [2] }

\label{highT}

In this section, we recall the Imry-Ma arguments \cite{ImryMa}
of Garel et al. \cite{garel} ,
 for the symmetric and dissymmetric cases respectively.
They are expected to be valid at high temperature.

\subsubsection{ Imry-Ma argument for the symmetric case  }

The Imry-Ma argument exposed in \cite{garel}
for the symmetric case ($q_0=0$) can be summarized as follows.
Assuming that the chain is localized around the interface
with typical blobs of length $l$ in each solvent, 
the typical energy gain per blob
scales as $\sqrt{\sigma l}$,
whereas the reduction of entropy per blob scales as $\ln l$
(in \cite{garel}, only powers of $l$ were considered
for the symmetric case,
and thus $\ln l$ was replaced by $l^0 \sim 1$, but since
this $\ln l$ dependence plays a crucial role in the dissymmetric case,
we prefer to keep this $\ln l$ dependence everywhere in this paper).
Neglecting prefactors of order 1, the optimization of the free-energy per monomer
\begin{eqnarray} 
f(l) \sim - \sqrt{ \frac{  \sigma }{ l }}  +  T \frac{\ln l}{l}
\end{eqnarray}
with respect to $l$ leads to
\begin{eqnarray} 
\frac{l}{(\ln l)^2} \sim \frac{ T^2 } { \sigma } 
\qquad . 
\end{eqnarray}
At high temperature, the scaling of the typical blob length $l(T)$
and of the free-energy $f(T)$ are therefore given by :
\begin{eqnarray} 
&& l(T) \sim \frac{ T^2 (\ln T)^2} { \sigma }  \qquad  ,\\
&& f(T) \sim  -\frac{\sigma}{T \ln T} \qquad 
\end{eqnarray}
and thus the heteropolymer remains localized at any finite temperature.

\subsubsection{ Imry-Ma argument for the dissymmetric case  }

The Imry-Ma argument exposed in \cite{garel}
for the dissymmetric case is actually much more subtle
than in the symmetric case, because to describe
correctly the blobs in the $(-)$ solvent, it is necessary
to consider the ``rare events" where the sum
of random variables $q_i$ of positive mean $\overline{q_i}=q_0>0$
happens to be negative enough in order
to make more favorable for the heteropolymer
to make an excursion in the $(-)$ solvent 
rather than to stay in the otherwise preferred $(+)$ solvent.
More precisely, the Imry-Ma argument of Reference \cite{garel} is as follows
in our notations : at high temperature, the heteropolymer is expected
to be mostly in the preferred $(+)$ solvent, except
when a blob of length $l^-$ in the $(-)$ solvent becomes energetically
favorable, with a blob energy $Q_-=-\sum_{i=j}^{j+l^-} q_i>0$ 
that is ``sufficient". The probability to have $\sum_{i=j}^{j+l^-} q_i=-Q^-$
is given by the Gaussian 
\begin{eqnarray} 
{\rm Prob}(Q^-)= \frac{1}{\sqrt{4 \pi \sigma l^-}} 
e^{-\frac{(Q^-+q_0 l^-)^2}{4 \sigma l^-} }
\qquad .
\end{eqnarray}
Thus the typical spacing $l^+$
between two such events behaves as the inverse of this small probability
\begin{eqnarray} 
l^+ \sim e^{
\frac{(Q^-+ q_0 l^-)^2}{4 \sigma l^-} }
\qquad .
\end{eqnarray}
As a consequence, in a configuration with blobs of order $(l^+,l^-)$
with $l^+ \gg l^- $,
the energy gain $Q^-$ due to one excursion 
in the $(-)$ solvent behaves as
\begin{eqnarray} 
Q^- \sim \sqrt{4 \sigma l^- \ln l^+ }-q_0 l^-
\qquad , 
\end{eqnarray}
whereas the loss of entropy due to this excursion is of order 
$\ln l^++\ln l^- \simeq \ln l^+$.
The free-energy difference per monomer between this localized
state and the delocalized state in the preferred solvent
can then be estimated as (neglecting prefactors of order 1) :
\begin{eqnarray} 
f(T,l^+,l^-)-f_{deloc}(T) 
\sim \frac{1}{l^+}  \left(- Q^- + T \ln l^+ \right)
\sim \frac{1}{l^+}  \left( q_0 l^- - \sqrt{4 \sigma l^- \ln l^+ }
+ T \ln l^+ \right)
\qquad .
\end{eqnarray}
The optimization with respect to $l^-$ leads to
\begin{eqnarray} 
\label{relationlmlp}
l^- \sim \frac{\sigma}{q_0^2} \ln l^+
\end{eqnarray}
and
\begin{eqnarray} \label{qmlog}
Q^- \sim \frac{\sigma}{q_0} \ln l^+
\qquad 
\end{eqnarray}
Thus both the energy gain $Q^-$ and the entropy cost
have the same $\ln l^+$ dependence!  
As a consequence, the free-energy difference factorizes into
\begin{eqnarray}
\label{relationflp} 
f(T,l^+)-f_{deloc}(T) 
\sim (T- T_c) \frac{\ln l_+}{l_+}
\qquad ,
\end{eqnarray}
where the parameter  
 \begin{eqnarray} 
T_c \sim \frac{\sigma}{q_0}
\end{eqnarray}
thus represents the critical temperature between the localized phase $T<T_c$
and the delocalized phase $T>T_c$.
Contrary to the symmetric case, 
the behavior of the free-energy with respect to the temperature
remains unknown, since the typical blob length $l^+$
depends upon the temperature T in a way that is not determined
by the present argument.
Still, the relations (\ref{relationlmlp}) and (\ref{relationflp})
between the typical blobs lengths $l^+$ and $l^-$ and the free-energy $f(T)$
represent non-trivial tests for any theory constructed to describe
 the heteropolymer problem.

\section{Definition of an Effective Thermodynamics}

\label{RG}

To go beyond the Imry-Ma arguments of previous section, one
needs to consider probability distributions
of blob lengths and blob energies.
In Appendix \ref{appendixImryMa}, we reproduce an extension
of the usual Imry-Ma argument 
which was introduced by Garel and Orland \cite{garelImryMa}, 
and which shows how to get the asymptotic behavior
of the probability distribution $P(l)$
of the blob length $l$ in the limit of small $l$.
Here, to study the probability distributions
of blob lengths and blob energies, we will construct via a 
real-space renormalization procedure the ``optimal" 
Imry-Ma domain structure associated with a given heteropolymer
at a given temperature.
We will then argue that the configurations
corresponding to this optimal Imry-Ma domain structure
dominate in the partition function
 asymptotically at high temperature.

\subsection{ Definition of a Real-Space Renormalization procedure
to construct the optimal Imry-Ma domain structure }

\begin{figure}[thb] 
\null
\vglue 1.0cm
\centerline{\epsfxsize=10.0 true cm \epsfbox{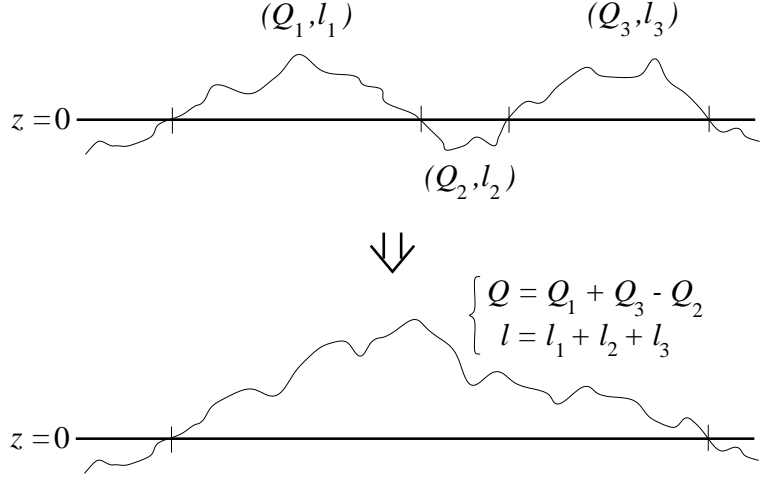}}
\vskip 1.0cm
\caption{
{\label{figRG} Illustration of the real-space renormalization procedure
: the flipping of the blob $(Q_2=\Gamma,l_2)$ surrounded 
by the two neighboring blobs $(Q_1,l_1)$ and $(Q_3,l_3)$
gives a new blob $(Q,l)$ with the rules (\ref{RGrules}). }} 
\end{figure}

At $T=0$, we have seen in Section \ref{ground}
that the chain is decomposed
 into blobs $\alpha$ containing $l_{\alpha}$ consecutive monomers
of charges of the same sign, and carrying an absolute charge $Q_{\alpha}$.
As $T$ grows from $T=0$, we consider the configurations of the chain obtained 
from the ground states structure by iteratively
flipping the blobs of smallest absolute charge $Q_{min} \equiv \Gamma$.
When we flip the blob $(Q_2=\Gamma,l_2)$ 
surrounded by the two neighboring blobs $(Q_1,l_1)$ and $(Q_3,l_3)$, 
we obtain a new blob of absolute charge $Q$
and length $l$ given by (see Figure \ref{figRG})
\begin{eqnarray}
&& Q=Q_1+Q_3-Q_2 \nonumber \\
&& l=l_1+l_2+l_3
\qquad .
\label{RGrules}
\end{eqnarray}

The renormalization procedure
corresponding to the rules (\ref{RGrules}) has already been
 extensively studied
in the context of the Random Transverse Field Ising Chain \cite{fisher},
where the role of the absolute charges $Q_i$ was played
by the logarithm of the random couplings and random fields
($\ln J_i$ and $\ln h_i$),
and in the context of Random walks in disordered 
1D environments \cite{RGSinai}
where the role of the absolute charges $Q_i$ was played
by the energy barriers $F_i$. As explained in details in Reference \cite{RGSinai},
the renormalization procedure constructs iteratively
the large scale extrema statistics of random walks,
where only barriers bigger than a given scale
$\Gamma$ are kept. Here the corresponding
random walk is simply given by the sum of the quenched
charges $\sum_{0}^i q_j$ as a function of the monomer index $i$,
and the renormalization procedure gives the ``best" blob structure,
given the constraint that only blobs of charges bigger than a given scale $\Gamma$ are kept.

To establish a relation between the renormalization
scale $\Gamma$ and the temperature $T$,
we now have to determine the conditions under which the flip of the blob $(Q_2,l_2)$
is favorable.
The cost in energy of the flip of the blob $(Q_2,l_2)$ is simply
\begin{eqnarray}
\Delta E^{flip} = 2 Q_2 \qquad ,
\end{eqnarray}
whereas the corresponding gain in entropy reads
\begin{eqnarray} \label{dsflip}
\Delta S^{flip} = \ln ({\cal M} (l_1+l_2+l_3)) 
-  \ln [{\cal M} (l_1) {\cal M} (l_2) {\cal M} (l_3) ]
\qquad ,
\end{eqnarray}
where ${\cal M} (l)$ represents the number of 1D random walks
of $l$ steps going from $z=0$ to $z=0$ in the presence of
an absorbing boundary at $z=0^-$.
As a consequence, the free-energy difference corresponding
to this flip reads
\begin{eqnarray}
\Delta F^{flip} = \Delta E^{flip} - T \Delta S^{flip}
= 2 Q_2 -  T \ln \left(\frac{ {\cal M} (l_1+l_2+l_3) }
{{\cal M} (l_1) {\cal M} (l_2) {\cal M} (l_3)} \right)
\qquad .
\label{flipcondition}
\end{eqnarray}

The optimal Imry-Ma domain structure corresponding to a given
realization of the disorder at a given temperature is thus
constructed as follows :
we iteratively flip the blobs of smallest absolute charge $Q$,
starting from the ground-states blob structure,
and continue as long as these flippings produce a decrease 
of the free-energy ($\Delta F^{flip} <0$).
We stop the procedure when the next flipping would correspond
to a raise of free-energy ($\Delta F^{flip} >0$).
We will define $\Gamma_{eq}(T)$ as the renormalization
scale $\Gamma$ where we have to stop the renormalization.
We now present a dynamical interpretation of the renormalization procedure.

\subsection{Dynamical interpretation of the renormalization procedure}

\label{dynamicsinter}

The renormalization procedure defined above
to construct the optimal blob structure at temperature $T$
can be given a perhaps more direct physical meaning if one considers
the dynamics at high temperature $T$ for $t>0$
starting from a zero-temperature initial condition at $t=0$,
i.e. after a quench to $T=0$ for $t<0$.
Indeed, the time necessary to flip a blob of absolute
charge $Q$ follows an Arrhenius law $t \sim t_0 e^{\beta Q}$
(where $t_0$ is a microscopic time scale).
The dynamics thus corresponds to the iterative flipping
of the blobs of smallest absolute charge remaining in the chain
described by the renormalization procedure, where
the renormalization scale $\Gamma$ now corresponds to time via 
\begin{eqnarray} 
\Gamma=T \ln \frac{t}{t_0} 
\end{eqnarray}
as in References \cite{RGletter} \cite{RGSinai}.
The dynamics takes place 
up to time $t_{eq}$ where equilibrium at temperature $T$ is reached
\begin{eqnarray} 
T \ln \frac{t_{eq}}{t_0}  = \Gamma_{eq}(T) \qquad .
\end{eqnarray}

\subsection{Discussion of the validity of the effective thermodynamics}

Let us define the effective partition function $Z^{\rm eff}_L(\beta;\{q_i\})$
as the sum over all the configurations that correspond to 
the disorder and temperature dependent 
optimal Imry-Ma domain structure constructed via
the renormalization procedure defined above.
The real partition function
$Z_L(\beta;\{q_i\})$ can be decomposed into
\begin{eqnarray} 
Z_L(\beta;\{q_i\}) = Z^{\rm eff}_L(\beta;\{q_i\})+
\hbox{ sum corresponding to other Imry-Ma domains structures}
\ .
\end{eqnarray}
The effective thermodynamics based on $Z^{\rm eff}_L(\beta;\{q_i\})$
is of course very approximate at low temperatures. 
However, at high temperatures,
the renormalization scale $\Gamma=Q_{min}$
will be large and the probability distributions
of absolute charges of the blobs will become infinitely broad \cite{fisher}.
More precisely in the symmetric case, the difference 
between the typical value $Q_{typ}$ of the absolute charges
and the minimum value $Q_{min}=\Gamma$ will also be
large and of order $\Gamma$ : $Q_{typ}-Q_{min} \sim \Gamma$.
As a consequence, we expect that all the configurations
that do not correspond approximatively 
to the optimal Imry-Ma domain structure
will be highly suppressed in the partition function,
in comparison with the ``optimal configurations".
The validity of the effective thermodynamics is thus 
based on the idea that, for a given heteropolymer
 at a given high temperature,
the large scale Imry-Ma domain structure
is essentially {\it unique}.

As explained in the Introduction,
in the other fields where disordered-dependent 
real space renormalization approaches have been used,
they have been able to reproduce the
non-trivial exact results obtained previously via other methods
(see References \cite{fisher} \cite{RGSinai} ).
These agreements with the Mc Coy-Wu exact results
in the field of disordered quantum spin chains \cite{fisher},
and with the Kesten-Golosov distribution in the field of
1D random walks in random media \cite{RGSinai} 
have provided the best evidence
for the asymptotic exactness
of the renormalization approach in these fields, and give
confidence in all its other novel predictions.
Here for the heteropolymer problem, 
we do not have at our disposal so precise exact analytical results
derived via other methods to ``test" unambiguously
the asymptotic exactness of the predictions of the renormalization approach.
As a consequence, we hope that in the future,
the explicit results of the renormalization approach
presented in this paper
will be tested precisely by numerical studies.
It would be also very interesting
to test more directly the idea of the
dominance of the optimal Imry-Ma domain structure
in the partition function, by comparing,
sample by sample, the typical configurations
of the heteropolymer at equilibrium at temperature $T$
with the optimal blob structure
obtained by the numerical implementation of the renormalization procedure
up to scale $\Gamma_{eq}(T)$.

The remainder of the paper is devoted to
the detailed study of the properties of the effective thermodynamics.
To proceed now with the concrete calculations of the renormalization
approach, it is again more convenient to discuss separately the symmetric case 
$q_0=0$ and the dissymmetric case $q_0 > 0$.

\section{Study of the symmetric case $q_0=0$}

\label{symmetric}

\subsection{Properties of the renormalization procedure}

It was shown in Reference \cite{fisher}
 that with the above renormalization procedure
 (\ref{RGrules}), the probability distribution
$P_{\Gamma}(Q,l)$ that a blob at renormalization scale $\Gamma$
has absolute charge $Q$ and length $l$ flows towards
a fixed point distribution $P^*(\eta,\lambda)$ in the rescaled variables
\begin{eqnarray} \label{rescaling}
\eta=\frac{Q-\Gamma}{\Gamma}   \qquad  \hbox{and}
  \qquad  \lambda= \frac{\sigma l}{\Gamma^2} \qquad ,
\end{eqnarray}
with $\sigma$ defined in (\ref{defq}),
and that this fixed point distribution $P^*(\eta,\lambda)$ reads in Laplace transform with respect to $\lambda$
\begin{eqnarray} \label{solu}
{\cal L}_{ \lambda \to s} \left(  P^*(\eta,\lambda) \right) \equiv
\int_0^{\infty} d \lambda e^{- s \lambda} P^*(\eta,\lambda) =
 \frac{\sqrt{s}}{\sinh \sqrt{s}  } \ 
e^{-\eta\sqrt{s}\coth \sqrt{s} }  
\qquad .
\end{eqnarray}
In particular, the distribution of the rescaled absolute charge $\eta$
is a simple exponential
\begin{eqnarray}
P^*(\eta)=\int_0^{\infty} d \lambda  P^*(\eta,\lambda)
=e^{-\eta} \qquad,
\end{eqnarray}
and the distribution of the rescaled length $\lambda$ of a blob reads
\begin{eqnarray}  \label{pdel}
&& P(\lambda)  = \int_0^{\infty} d \eta  P^*(\eta,\lambda)
= {\cal L}^{-1}_{s \to \lambda } \left( \frac{1}{ \cosh(\sqrt{s})}\right)
= \int_{-i \infty}^{+i\infty} 
\frac{ds}{2 i \pi} \frac{e^{s \lambda}}{ \cosh(\sqrt{s})} 
\nonumber \\
&&= \sum_{n = -\infty}^{\infty} \left(n+\frac{1}{2}\right)
\pi (-1)^n e^{- \pi^2 \lambda \left(n+\frac{1}{2}\right)^2}
= \frac{1}{\sqrt \pi \lambda^{3/2}}
\sum_{m = -\infty}^{\infty} (-1)^m (m+\frac{1}{2})
 e^{-  \frac{1}{\lambda} (m+\frac{1}{2})^2}
\qquad .
\end{eqnarray}
We also introduce the notation $P_{\rm flip}(\lambda)$
for the distribution of the bonds which are 
about to be decimated ( i.e. having $\eta=0$ )
\begin{eqnarray}  \label{pdeci}
&& P_{\rm flip}(\lambda)  
={\cal L}^{-1}_{s \to \lambda } \left( \frac{\sqrt{s}}{ \sinh(\sqrt{s})}\right) \nonumber \\
&&= \pi^2 \sum_{n = -\infty}^{\infty} 
 (-1)^{n-1} n^2 e^{- \pi^2 n^2 \lambda }
= \frac{1}{ 2 {\sqrt \pi} \lambda^{3/2}}
\sum_{m = -\infty}^{\infty} \left[ \frac{2}{\lambda} (m+\frac{1}{2})^2 -1
\right]  e^{-  \frac{1}{\lambda} (m+\frac{1}{2})^2}
\qquad .
\end{eqnarray}

\subsection{Relation between temperature and renormalization scale}

We now examine the condition $\Delta F^{flip}<0$, where
the free-energy difference $\Delta F^{flip}$ due to a blob flip was given in 
Equation (\ref{flipcondition}).
For large $l$, the number ${\cal M} (l)$ of 1D random walks
of $l$ steps going from $z=0$ to $z=0$ in the presence of
an absorbing boundary at $z=0^-$ behaves as
\begin{eqnarray} \label{mlasymp}
{\cal M} (l) \simeq \kappa \frac{\mu^l}{l^{3/2}}
\qquad ,
\end{eqnarray}
where $\kappa$ is a constant, and where $\ln \mu$ characterizes
the entropy of the free random walk
(for instance $\mu=2$ for the 1D lattice).
In the effective thermodynamics at large 
renormalization scale $\Gamma$,
 the free-energy difference of a blob flip then reads
\begin{eqnarray}
\Delta F^{flip} =
 2 Q_2 -  T \ln \left(\frac{ (l_1 l_2 l_3)^{3/2} }
{ \kappa^2 (l_1 +l_2 +l_3)^{3/2} } \right)
\qquad .
\end{eqnarray}
Using the rescaled variables $\lambda_i=\frac{\sigma l_i}{\Gamma^2}$, 
we obtain the average
\begin{eqnarray}
\overline{\Delta F^{flip} } =
 2 \Gamma -  3 T \ln \frac{\Gamma^2}{\sigma} 
+T ( \frac{3}{2}  K + \ln \kappa^2)
\qquad ,
\end{eqnarray}
where $K$ is a pure numerical constant that can be obtained from
the distributions (\ref{pdel}) and (\ref{pdeci}) as
\begin{eqnarray}
K && = \int_0^{\infty} d \lambda_1 P^*(\lambda_1)
\int_0^{\infty} d \lambda_2 P^*_{\rm flip}(\lambda_2)
\int_0^{\infty} d \lambda_3 P^*(\lambda_3) 
\ln (\lambda_1+\lambda_2+\lambda_3) 
\nonumber \\
&& - 2 \int_0^{\infty} d \lambda P^*(\lambda) \ln \lambda
- \int_0^{\infty} d \lambda P^*_{\rm flip}(\lambda) \ln \lambda
\qquad .
\end{eqnarray}

The scale $\Gamma_{eq}(T)$ where the renormalization procedure has to be stopped
is the solution of $\Delta F^{flip}(\Gamma)=0$ and $\partial_{\Gamma}
\Delta F^{flip}(\Gamma)>0$. For large $\Gamma$, and thus for large
temperature, we find that $\Gamma_{eq}(T)$
is implicitly defined as the inverse of the function $T(\Gamma_{eq})$
given by
\begin{eqnarray} \label{gammaeqt}
T(\Gamma_{eq}) \simeq \frac{  \Gamma_{eq}}{ 3 \ln \Gamma_{eq}}
\qquad .
\end{eqnarray}

As a consequence, for any arbitrary large temperature $T$,
we find that the renormalization procedure has to be stopped at 
the {\it finite} large renormalization scale $\Gamma_{eq}(T)$.
The polymer is thus always localized around the interface,
with blobs of typical length $l^{\rm blob}(T)$
behaving at high temperature as
\begin{eqnarray}
l_{ blob}(T) \sim \frac{\Gamma_{eq}^2}{\sigma} 
\sim \frac{T^2 (\ln T)^2 }{\sigma}
\qquad .
\end{eqnarray}
In this regime, the typical distance $z(T)$
to the interface behaves as
\begin{eqnarray}
z(T) \sim \sqrt{ D l_{ blob}(T)} \sim \sqrt{\frac{D}{\sigma}}
\Gamma_{eq} \sim \sqrt{\frac{D}{\sigma}} T (\ln T)
\qquad .
\end{eqnarray}
These scaling behaviors are in agreement
with the Imry-Ma argument of Reference \cite{garel} (see Section \ref{highT},
where the logarithmic correction is found to be present
in the Imry-Ma argument) and with 
the exact free energy bounds of Reference \cite{maritanbounds}.
The Replica 
variational approaches of References \cite{stepanow} \cite{maritanreplica}
give the same scaling behaviors without the logarithmic correction.

We now study more precisely various physical quantities
for which the present renormalization approach yields 
explicit predictions. 

\subsection{ Thermodynamic quantities }

\label{thermosym}

The energy of a blob is simply given by minus its absolute charge $(-Q)$,
and thus the energy per monomer $e(T)$ of the chain
in the thermodynamic limit
can be obtained through a decomposition into blobs indexed by $\alpha$ as
\begin{eqnarray} \label{esymm}
e(T)= - \frac{\sum_{\alpha} Q_{\alpha} }
{\sum_{\alpha} l_{\alpha}} = -
\frac{\int_{Q,l} Q P_{\Gamma_{eq}(T)}(Q,l)  }{ \int_{Q,l} l P_{\Gamma_{eq}(T)}(Q,l)}
\qquad .
\end{eqnarray}
with $\Gamma_{eq}(T)$ defined in (\ref{gammaeqt}).
Using the fixed point solution of Equations (\ref{rescaling}, \ref{solu}),
 we obtain at large temperature
\begin{eqnarray}
e(T)\simeq - \frac{4 \sigma}{\Gamma_{eq}(T)}
 \sim - \frac{4 \sigma}{ 3 T \ln T}
\qquad .
\end{eqnarray}

The entropy associated with the $z$ direction
for a blob of length $l$ is given by
$\ln ({\cal M} (l))$, where ${\cal M} (l)$ 
behaves as in Equation (\ref{mlasymp}) in the large $l$ regime.
As a consequence, the entropy per monomer $s(T)$
is given in the thermodynamic limit by 
\begin{eqnarray} \label{ssymm}
s(T)= \frac{\sum_{\alpha} [l_{\alpha} \ln \mu 
+\ln \kappa -\frac{3}{2} \ln l_{\alpha} ] }
{\sum_{\alpha} l_{\alpha}} =
\frac{\int_{Q,l} [l \ln \mu 
+\ln \kappa -\frac{3}{2} \ln l ] P_{\Gamma_{eq}(T)}(Q,l)  }{ \int_{Q,l} l P_{\Gamma_{eq}(T)}(Q,l)}
\qquad .
\end{eqnarray}
Using again the fixed point solution (\ref{rescaling} , \ref{solu}), we have
at large temperature 
\begin{eqnarray}
s(T) && \simeq \ln \mu - \frac{6 \sigma \ln(\Gamma_{eq}(T))}{\Gamma^2_{eq}(T)}+
\frac{\sigma}{\Gamma^2_{eq}(T) }
\left( 3 \ln \sigma + 2 \ln \kappa
-3 \int_0^{\infty} d \lambda P^*(\lambda) \ln \lambda  \right) 
\nonumber \\
&& \simeq \ln \mu - \frac{2 \sigma }{ 3 T^2 \ln T}
 + O\left(\frac{1}{T^2 (\ln T)^2} \right)
\qquad .
\end{eqnarray}

Finally, we obtain the free-energy per monomer $f(T)$ of the chain
at large temperature 
\begin{eqnarray} \label{freesym}
f(T) && \simeq  -T \ln \mu -\frac{2 \sigma}{\Gamma_{eq}(T)}+
\frac{\sigma}{\Gamma_{eq}(T) \ln \Gamma_{eq}(T)}
\left( \int_0^{\infty} d \lambda P^*(\lambda) \ln \lambda -\ln \sigma - \frac{2}{3} \ln \kappa \right) 
\nonumber \\
&& \simeq  -T \ln \mu -\frac{2 \sigma}{ 3T \ln T}+ 
O\left(\frac{1}{T (\ln T)^2} \right)
\qquad .
\end{eqnarray}
The dominant behavior depending on the disorder
parameter $\sigma$ is in agreement with the Imry-Ma argument
of Reference \cite{garel} as explained in Section \ref{highT}.

\subsection{ Distribution of the blob lengths}

In the renormalization approach, the lengths of the different blobs are 
independent identical random variables,
and the rescaled blob length defined as
\begin{eqnarray}
\lambda=\frac{\sigma l_{blob} }{\Gamma_{eq}^2(T)}
\end{eqnarray}
is asymptotically distributed at high temperature
with the distribution $P^*(\lambda)$
given in Equation (\ref{pdel}). 
In particular, its asymptotic behaviors for large and small
 $\lambda$ are respectively given by
\begin{eqnarray}  \label{pdelinfty}
 P(\lambda)  \opsimeq_{\lambda \to \infty} 
\pi  e^{- \frac{\pi^2}{4} \lambda } 
\end{eqnarray}
\begin{eqnarray}  \label{pdelzero}
 P(\lambda)  \opsimeq_{\lambda \to 0} 
 \frac{1}{\sqrt \pi \lambda^{3/2}}
 e^{-  \frac{1}{ 4 \lambda} }
\qquad .
\end{eqnarray}

As noted by Garel and Orland \cite{garelImryMa},
the small $\lambda$ behavior found in Equation (\ref{pdelzero})
for the distribution $P(\lambda)$ can be obtained
via an extension of the usual Imry-Ma argument,
as explained in the Appendix (\ref{appendixImryMa}). 

\subsection{ Density $\rho(z)$ of polymer at a distance $z$ from the interface}

We now study the density $\rho(z)$ of polymer at a distance $z$ 
from the interface as follows.
We first need to introduce the probability
$S_{\Gamma} (l,x)$ that a given point
of the chain belongs at renormalization scale $\Gamma$ to a blob
of length $l$ and is at distances $(x,l-x)$ from the ends of the blob.
It reads
\begin{eqnarray}
S_{\Gamma} (l,x) = \frac{P_{\Gamma}(l)}
{\int_0^{\infty} dl \ l \ P_{\Gamma}(l)} \theta(x) \theta(l-x)
\qquad ,
\end{eqnarray}
and is normalized to 
$\int_0^{\infty} dl \int_0^l dx S_{\Gamma} (l,x)=1$.
We then need the probability $B_{x,l}(z)$ 
to be at height $z$ at ``time" $x$ for a Brownian motion 
of diffusion coefficient $D$ starting at $z=0$ at $x=0$
and finishing at $z=0$ at $x=l$ in the presence of a reflexive 
boundary at $z=0$ 
\begin{eqnarray}
B_{x,l}(z) = \frac{2}{\sqrt \pi} \left( \frac{l}{ 2 D x (l-x)} \right)^{1/2} 
e^{- z^2 \frac{l}{ 2 D x (l-x)} }
\qquad .
\end{eqnarray}
It is normalized to $\int_0^{\infty} dz B_{x,l}(z)=1$.
The probability $\rho(z)$ that a given point of the polymer
is at a distance $z$ from the interface (normalized to $\int_0^{\infty} dz \rho(z)=1$) can be now expressed within the renormalization picture as
\begin{eqnarray}  \label{densitysym}
 \rho(z) 
&& = \int_0^{\infty} dl \int_0^l dx S_{\Gamma_{eq}(T)} (l,x) B_{x,l}(z) 
\nonumber \\
&& =  \frac{1}
{\int_0^{\infty} dl \ l \ P_{\Gamma}(l)}
\int_0^{\infty} dl P_{\Gamma}(l) \ 
4 \sqrt{\frac{l}{ 2 D}} \int_{\frac{2z}{\sqrt{ 2 D l}}}^{\infty} du e^{-u^2}
\qquad .
\end{eqnarray}
Introducing again the rescaled variable $\lambda=\frac{\sigma l}{\Gamma^2}$,
we thus obtain the scaling form
\begin{eqnarray}
\rho(z) dz = R(Z) dZ 
\qquad ,
\end{eqnarray}
where the rescaled distance $Z$ to the interface, defined as
\begin{eqnarray}
 Z=  \sqrt{\frac{2 \sigma }{D}} \frac{z}{\Gamma_{eq}(T)}
\qquad ,
\end{eqnarray}
is distributed with
the following scaling function  
\begin{eqnarray}
R(Z)= 4 \int_0^{\infty} d \lambda P^*(\lambda) \sqrt{\lambda}
\int_{\frac{Z}{\sqrt{\lambda}}}^{\infty} du e^{- u^2}
\qquad .
\end{eqnarray}
where $P^*(\lambda)$ is the fixed point
distribution of Equation (\ref{pdel}).
Deriving with respect to $Z$ yields
\begin{eqnarray}
R'(Z)  = - 4 \int_0^{\infty} d \lambda P^*(\lambda) 
  e^{- \frac{Z^2}{\lambda} } 
 = -16 Z \sum_{n=0}^{\infty} (-1)^n K_1 \left[ (2n+1)\pi Z \right] 
\end{eqnarray}
in terms of the Bessel function $K_1(y)$.

In particular, the asymptotic behaviors at $Z \to 0$ and $Z \to \infty$
are given respectively by
\begin{eqnarray}
&& R(Z)= R(0) -4 Z +O(Z^2) \qquad , \\
&& R(0)=2 \sqrt{\pi} \int_0^{\infty} d \lambda P^*(\lambda) \sqrt{\lambda}
= \frac{8}{\pi} \sum_{n=0}^{\infty} \frac{(-1)^n}{(2n+1)^2}
\qquad ,
\end{eqnarray}
and
\begin{eqnarray}
R(Z) \opsimeq_{Z \to \infty}  \frac{8}{\pi} \sqrt{2 Z} e^{-\pi Z} \qquad .
\end{eqnarray}

We may also compute the moments
\begin{eqnarray}
\int_0^{\infty} dZ Z^k R(Z)
&& = 2 \frac{\Gamma(1+\frac{k}{2})}{k+1} 
\int_0^{\infty} d \lambda P^*(\lambda) \lambda^{1+\frac{k}{2}} 
\nonumber \\
&& = 4 \frac{ \Gamma(1+\frac{k}{2}) \Gamma(2+\frac{k}{2})}
{(k+1) \pi^{3+k}} 
\sum_{n=0}^{\infty} \frac{(-1)^n}{ (n+\frac{1}{2}) ^{3+k} } 
\qquad .
\end{eqnarray}

In particular, the mean value of the square distance to the interface
is given at high temperature  by
\begin{eqnarray}
\overline{<z^2>}  \simeq \Gamma_{eq}^2(T) \frac{D}{2 \sigma} 
\int_0^{\infty} dZ Z^2 R(Z) = \frac{ 5 D}{ 36 \sigma} \Gamma_{eq}^2(T) 
\simeq \frac{ 5 D}{ 4 \sigma} T^2 (\ln T)^2
\qquad .
\end{eqnarray}

\subsection{ Density $\rho_{a,b}(z,z')$ 
for two thermal copies of the same chain }

To compare our approach with the Replica Gaussian Variational 
description of Reference \cite{maritanreplica}, we now compute the probability
 $\rho_{a,b}(z,z')$ that a given monomer of a given polymer chain
is at distance $z$ from the interface in a configuration (a)
and at a distance $z'$ from the interface in a configuration (b)
(note that in the effective thermodynamics, 
we have ${\rm sgn}(z)={\rm sgn}(z')$).
Using the same notations as in the previous section, the joint distribution
of $z$ and $z'$ can be obtained as
\begin{eqnarray}
 \rho_{a,b}(z,z') 
&& = \int_0^{\infty} dl \int_0^l dx S_{\Gamma_{eq}(T)} (l,x) B_{x,l}(z)
 B_{x,l}(z') \nonumber \\
&& =  \frac{1}
{\int_0^{\infty} dl \ l \ P_{\Gamma}(l)}
\int_0^{\infty} dl P_{\Gamma}(l) \ 
\frac{4}{\pi D}  e^{-\frac{z^2+z'^2}{D l^2}}
K_0 \left( \frac{z^2+z'^2}{D l^2} \right)
\qquad .
\end{eqnarray}
Using again the rescaled variable $\lambda=\frac{\sigma l}{\Gamma^2}$,
we obtain the scaling form
\begin{eqnarray}
\rho_{a,b}(z,z') dz dz' = R_{a,b} (Z,Z') dZ dZ' \qquad \qquad \hbox{with}
\qquad Z=  \sqrt{\frac{2 \sigma }{D}} \frac{z}{\Gamma_{eq}(T)}
\qquad .
\end{eqnarray}
The scaling function $R_{a,b} (Z,Z')$ reads in terms of the fixed point
distribution $P^*(\lambda)$ of Equation (\ref{pdel})
\begin{eqnarray}
R_{a,b} (Z,Z')= \frac{4}{\pi} \int_0^{\infty} d \lambda P^*(\lambda) 
e^{-\frac{Z^2+Z'^2}{2 \lambda}} K_0 \left(\frac{Z^2+Z'^2}{2 \lambda} \right)
\qquad .
\end{eqnarray}
It is interesting to note that it is actually a function of
the single variable $(Z^2+Z'^2)$.

The probability that a given monomer is
 at the same $z$-coordinate
in the two configurations (a) and (b) thus
behaves at high temperature as
\begin{eqnarray}
\int_0^{\infty} dz \rho_{a,b}(z,z)  = 
\sqrt{\frac{2 \sigma }{D}} \frac{1}{\Gamma_{eq}(T)}
\int_0^{\infty} dZ R_{a,b} (Z,Z) \opsim_{T \to \infty} \frac{1}{T \ln T}
\qquad ,
\end{eqnarray}
as compared to the decay as $(1/T)$ found in Reference \cite{maritanreplica}
for the Replica symmetric solution.

\subsection{Solvent-solvent correlation function 
$\overline{< {\rm sgn} (z(s)) {\rm sgn} (z(s')) > }$}

Within the renormalization picture, the correlation function 
$\overline{< {\rm sgn} (z(s)) {\rm sgn} (z(s')) > }$ 
of the solvents seen by two monomers at distance $(s-s')$
along the polymer chain turns out
to correspond exactly to the spin correlation function
$\overline{\langle S_0(t)S_{x}(t) \rangle}$ of 
the random field Ising model in the Glauber
dynamics starting from a random initial condition
computed in References \cite{RGletter} \cite{RGRFIM}.
We thus only quote the result

\begin{eqnarray} 
 \overline{< {\rm sgn} (z(s)) {\rm sgn} (z(s')) > }
&&= {\cal L}^{-1}_{p \to X= \sigma \frac{\vert s-s' \vert}{\Gamma^2} }\left[ 
\frac{1}{p}-
\frac{4}{p^2} \rm{tanh}^2 \left( \frac{\sqrt p}{2} \right)\right]  
\nonumber \\
&& = \sum_{n=-\infty }^{\infty } 
\frac{48 (2n+1)^2\pi^2 + 32 \sigma \frac{\vert  s-s' \vert}{\Gamma^2 }}{(2n+1)^4\pi^4} 
e^{-(2n+1)^2\pi^2 \sigma \frac{\vert  s-s' \vert}{\Gamma^2}} 
\qquad ,
\end{eqnarray}
where $\Gamma=\Gamma_{eq}(T)$. In particular, the correlation 
length $\xi(T)$ associated with this correlation function
is given by the decay of the dominant exponential at large separation 
$\vert  s-s' \vert$
\begin{eqnarray} 
\xi(T) = \frac{\Gamma^2_{eq}(T)}{\pi^2 \sigma} 
\simeq \frac{9 T^2 (\ln T)^2}{\pi^2 \sigma}
\qquad .
\end{eqnarray}

\section{Study the biased case $q_0>0$}

\label{biased}

\subsection{Properties of the renormalization procedure}

In the biased case, it is necessary to introduce
two probability distributions $P_{\Gamma}^{+}(Q,l)$
and $P_{\Gamma}^{-}(Q,l)$ representing the probabilities
for a blob at scale $\Gamma$ in the $(\pm$) solvent
to have absolute charge $Q$ and length $l$. It
 was shown in Reference \cite{fisher} that with the renormalization procedure
(\ref{RGrules}), these probability distributions
 flow at large $\Gamma$ towards
\begin{eqnarray} 
\label{solu-biased}
 P_{\Gamma}^{\pm}(Q,l)   & = &
{\cal L}^{-1}_{p \to l} \left( U_{\Gamma}^{\pm}(p ) e^{- (Q-\Gamma) 
u_{\Gamma}^{\pm}(p)} \right) \\
u_{\Gamma}^{\pm}(p) & = & \Delta(p) \coth{[\Gamma \Delta(p)]} 
\mp \delta  \nonumber \\
U_{\Gamma}^{\pm}(p)  & = & 
\frac{\Delta(p)} {\sinh{[\Gamma \Delta(p)]} } e^{\mp \delta \Gamma} 
\nonumber \\
\Delta(p) & = & \sqrt{\delta^2+\frac{p}{\sigma}}
\qquad , \nonumber 
\end{eqnarray}
where $\sigma$ and $\delta$ have been defined
in Equations (\ref{defq}) and (\ref{defdelta}) in terms
of the initial charge distribution $C(q)$.
The solutions (\ref{solu-biased}) are valid
 in the scaling  regime of large $\Gamma$, small $\delta$ and
small $p$ with $\delta \Gamma$ fixed and $p \Gamma^2$ fixed \cite{fisher}.

In particular, the distributions of the absolute charge $Q \geq \Gamma$
are simple exponentials :
\begin{eqnarray} \label{distriQ}
&& P^{+}_{\Gamma}(Q) = u_{\Gamma}^+  e^{- (Q-\Gamma) u_{\Gamma}^+}
\qquad \qquad \hbox{ with} \qquad 
 u_{\Gamma}^+= \frac{2 \delta}{e^{2 \Gamma \delta} - 1}
\qquad , \\
&&  P^{-}_{\Gamma}(Q) = u_{\Gamma}^-  e^{- (Q-\Gamma) u_{\Gamma}^-}
\qquad \qquad \hbox{ with} \qquad 
 u_{\Gamma}^-= \frac{2 \delta}{1 - e^{-2 \Gamma \delta}}
\qquad ,
\end{eqnarray}
whereas the distributions of the length of blobs 
have the form of an infinite series of exponentials
\cite{RGSinai}
\begin{eqnarray}  \label{pdelpm}
 P^{\pm}_{\Gamma}(l)  
&& = {\cal L}^{-1}_{p \to l } 
\left( \frac{ \Delta(p) e^{\mp \delta \Gamma}}
{\Delta(p) \cosh{[\Gamma \Delta(p)]} 
\mp \delta \sinh{[\Gamma \Delta(p)]} }\right) 
\nonumber \\
&&= \frac{\sigma}{\Gamma^2} \sum_{n=0}^{\infty}
J_n^{\pm}(\gamma=\delta \Gamma) e^{- \frac{\sigma l}{\Gamma^2}
s_n^{\pm}(\gamma=\delta \Gamma) }
\qquad ,
\end{eqnarray}
where the functions
$s_n^{\pm}(\gamma)$ and $J_n^{\pm}(\gamma)$
are defined in terms of the roots 
$\alpha^{\pm}_{n}(\gamma)$ ($n=0,1,...$) of the equation
\begin{eqnarray}
\alpha^{\pm}_{n}(\gamma) {\rm cotan} (\alpha^{\pm}_{n}(\gamma)) = \pm \gamma
\qquad \hbox{with} \qquad  n \pi < \alpha^{\pm}_{n}(\gamma) < (n+1) \pi
\qquad .
\label{anlaplace}
\end{eqnarray}
For $\gamma > 1$, the root $\alpha^{+}_{0}(\gamma)$ does not exist,
but is replaced by the positive root $\tilde{\alpha}^{+}_{0}(\gamma)$
of the equation $\tilde{\alpha}^{+}_{0}(\gamma) \coth(\tilde{\alpha}^{+}_{0}(\gamma)) =  \gamma$.
In terms of these roots, we have
\begin{eqnarray} \label{snlaplace}
&& s_n^{\pm}(\gamma)=\gamma^2+ (\alpha^{\pm}_{n}(\gamma))^2
\qquad , \\
&& J_n^{\pm}(\gamma)=
\frac{ 2 (-1)^{n} (\alpha^{\pm}_{n}(\gamma))^2 
\sqrt{\gamma^2+ (\alpha^{\pm}_{n}(\gamma))^2} e^{\mp \gamma} }
{  \gamma^2
+ (\alpha^{\pm}_{n}(\gamma))^2 \mp \gamma }
\qquad ,
\end{eqnarray}
except for the $(+)$ $n=0$ term in the domain $\gamma>1$ for which
\begin{eqnarray}
&& s_0^{+}(\gamma>1)=\gamma^2- (\tilde{\alpha}^{+}_{0}(\gamma))^2  \\
&& J_0^{+}(\gamma>1)=
\frac{ 2  (\tilde{\alpha}^{+}_{0}(\gamma))^2 
\sqrt{\gamma^2- (\tilde{\alpha}^{+}_{0}(\gamma))^2} e^{- \gamma} }
{ \gamma
+(\tilde{\alpha}^{+}_{0}(\gamma))^2-\gamma^2 }
\qquad .
\end{eqnarray}

The mean lengths of the blobs in the domains $(\pm)$ have simple
expressions
\begin{eqnarray} \label{lengthpm}
&&\overline{l^{+}_{\Gamma}} = \frac{1}{4 \delta^2 \sigma } (e^{2\delta \Gamma} 
-2  \delta \Gamma -1 ) 
\qquad , \\
&&\overline{l^{-}_{\Gamma}} = \frac{1}{4 \delta^2 \sigma } ( 2\delta \Gamma- 1+
e^{- 2 \delta \Gamma}  )
\qquad ,
\end{eqnarray}
but the mean values of $\ln l^{\pm}$ are unfortunately much more complicated :
\begin{eqnarray}
\overline {\ln l^{\pm} }
=\int_0^{\infty} dl (\ln l) P_{\Gamma}^{\pm} (l)
=\sum_{n=0}^{\infty} \frac{J_n^{\pm}(\gamma)} {s_n^{\pm}(\gamma) } 
\ln \left( \frac{1} {s_n^{\pm}(\gamma) } \right) 
+\ln \left( \frac{\Gamma^2}{\sigma} \right) -C_{\rm Euler}
\qquad ,
\end{eqnarray}
where $C_{\rm Euler}$ is the Euler constant.

\subsection{Relation between temperature and renormalization scale}

\label{relationbiased}

Using again Equation (\ref{mlasymp}), the averaged free-energy difference 
corresponding to the flip
of a blob out of the domain $(\pm)$ reads
\begin{eqnarray}
 \overline{ \Delta F^{flip(\pm)} }
= 2 \Gamma -  T \overline{ \ln \left(
\frac{ (l_1^\mp l_2^\pm l_3^\mp)^{3/2} }
      {(l_1^\mp +l_2^\pm +l_3^\mp)^{3/2} \kappa^2 } \right) } 
\qquad ,
\end{eqnarray}
which yields to a
 quite complicated expression for arbitrary $\gamma=\delta \Gamma$.

In the following, we consider the case of large $\gamma \gg 1$, 
where the distribution $P_{\Gamma}^+$ is dominated by the $n=0$ term
\cite{RGSinai}
\begin{eqnarray} \label{asymp+}
P_{\Gamma}^+(l) \simeq a^+(\gamma) 
e^{- a^+(\gamma) l } \qquad \hbox{with}  \qquad
a^+(\gamma) \simeq 4 \delta^2 \sigma  e^{-2\gamma}
\qquad ,
\end{eqnarray}
leading to
\begin{eqnarray}
\overline {\ln l^{+} } \simeq \ln \frac{1}{a^+(\gamma)} -C_{\rm Euler}
= 2\gamma + \ln \frac{1}{4 \delta^2 \sigma}-C_{\rm Euler}
\qquad .
\end{eqnarray}
In this regime, the probabilities to have blobs with absolute charge
$Q=\Gamma$ in the $(\pm)$ domain read respectively (\ref{distriQ})
\begin{eqnarray} 
&& P^{+}(Q=\Gamma) = u_{\Gamma}^+ \simeq 2 \delta e^{-2 \Gamma \delta} 
\qquad , \\
&&  P^{-}(Q=\Gamma) = u_{\Gamma}^-  \simeq 2 \delta
\qquad ,
\end{eqnarray}
and thus the renormalization consists asymptotically in flipping only blobs
out of the $(-)$ domain. We thus now concentrate on the free-energy difference corresponding to the flip
of a $(-)$ blob, which can be estimated as follows
\begin{eqnarray}
 \overline{ \Delta F^{flip(-)} }
&&=
 2 Q_2 -  T \overline{ \ln \left(
\frac{ (l_1^+ l_2^- l_3^+)^{3/2} }
      {(l_1^+ +l_2^- +l_3^+)^{3/2} \kappa^2  } 
\right) }  \nonumber \\
&& \simeq 2 \Gamma - \frac{3}{2} T 
\left[   2 \overline{\ln (l_1^+)} - \overline{\ln (l_1^+ + l_3^+)} \right] 
\nonumber \\
&& \simeq 2 \Gamma - \frac{3}{2} T \left[ 
2\delta \Gamma + \ln \frac{1}{4 \delta^2 \sigma}-C_{\rm Euler} -1\right] \qquad ,
\end{eqnarray}
so that finally, in the regime $\delta \to 0$, $\Gamma \to \infty$, 
with $\delta \Gamma \gg 1$ fixed, and large temperatures, we have
\begin{eqnarray}
 \overline{ \Delta F^{flip(-)} }
\simeq    \Gamma ( 2- 3 \delta T ) - \frac{3}{2} T  
 \ln \frac{1}{ \delta^2 \sigma}
\qquad .
\end{eqnarray}

Since the renormalization procedure has to be performed as long as 
$ \overline{ \Delta F^{flip(-)} } <0$, we obtain a
transition at the temperature 
\begin{eqnarray}
T_c=\frac{2}{3 \delta} \qquad ,
\end{eqnarray}
where $\delta$ is the parameter characterizing the 
distribution of quenched charges (\ref{defdelta}).
For $T<T_c$, the renormalization procedure has to be stopped at scale
$\Gamma_{eq}(T)$ given by
\begin{eqnarray} \label{geqd}
\Gamma_{eq} (T)=  \frac{ 3 T \ln \frac{1}{ \delta^2 \sigma} }{ 2 ( 2- 3 \delta T )}
=  \frac{ 3 T \ln \frac{1}{ \delta^2 \sigma} }{ 4 ( 1-\frac{T}{T_c} )} 
\qquad .
\end{eqnarray}
At the temperature $T=T_c$, the renormalization scale $\Gamma_{eq} (T)$
diverges : this corresponds to the delocalization transition
found previously in References \cite{garel} \cite{sommer} \cite{stepanow}  \cite{maritanreplica} \cite{maritanbounds} \cite{bolthausen}.

\subsection{ Comparison with previous results for the critical temperature}

\subsubsection{Case of Gaussian distribution for the quenched charges}

For the case where the initial distribution
of charges is Gaussian (\ref{gauss}), we get
\begin{eqnarray}
T_c=\frac{2}{3 \delta}= \frac{4 \sigma}{3 q_0}
\qquad ,
\end{eqnarray}
 in agreement with the scaling obtained 
previously \cite{garel}
 \cite{stepanow} \cite{maritanreplica}
and with the bounds established in Reference \cite{maritanbounds},
which read in terms of our notations
\begin{eqnarray}
\frac{2 \sigma}{\sqrt{\pi} \ln 2}
<T_c < \frac{2 \sigma}{q_0}
\qquad .
\end{eqnarray}
We note moreover that the value of the transition temperature obtained here
coincides with the critical temperature obtained
by Stepanow et al. \cite{stepanow} and is lower
than the critical temperature obtained
by Trovato et al. \cite{maritanreplica},
since their results read respectively in our notations :
\begin{eqnarray}
&& T_c^{{\rm Stepanow \ \ et } \ al.}= \frac{4 \sigma}{3 q_0}
\qquad , \\
&& T_c^{{\rm Trovato \ \  et } \ al.}= \frac{2 \sigma}{q_0}
\qquad .
\end{eqnarray}

\subsubsection{Case of binary distribution for the quenched charges}

For the case where the initial distribution
of charges is a binary distribution,
we obtain that $T_c$ is the solution of the equation (\ref{deltabinary})
\begin{eqnarray} \label{tcbinary}
c e^{- \frac{4}{3 T_c} q_+} +(1-c) e^{+\frac{4}{3 T_c} q_-} =1
\qquad .
\end{eqnarray}
In particular for the case $q_-=q_+=q_1$ (\ref{binarypm}), we obtain
\begin{eqnarray} 
 T_c=\frac{ 4 q_1}{3 \ln \left( \frac{c}{1-c} \right)}
\qquad ,
\end{eqnarray}
which is consistent with the bounds given in Reference \cite{maritanbounds}
 reading in our notations :
\begin{eqnarray} 
 T_c < \frac{ 4 q_1}{ \ln \left( \frac{c}{1-c} \right)}
\qquad .
\end{eqnarray}
We may also compare with the bounds obtained in Reference \cite{bolthausen}
concerning the binary case with $c=\frac{1}{2}$, $q_+=1+h$ and $q_-=1-h$
where $0<h<1$ for which we have :
\begin{eqnarray} 
q_0=h \qquad , \qquad 2 \sigma=1 \qquad , \qquad  e^{- 2 \delta h} \cosh \delta =1
\qquad .
\end{eqnarray}
Our renormalization approach yields thus for this case the following behavior
for the critical line $h_c(T)$ at high temperature
 \begin{eqnarray} 
h_c(T)= \left(\frac{1}{2 \delta} \ln \cosh \delta \right)\vert_{\delta
=\frac{2}{3 T}} = \frac{3 T_c}{4} \ln \cosh \left(\frac{2}{3 T_c} \right)
= \frac{1}{6 T} - \frac{1}{81 T^3} + O(\frac{1}{T^5})
\qquad ,
\end{eqnarray}
which satisfies the bounds established in Reference \cite{bolthausen}
\begin{eqnarray} 
0 < \lim_{T \to \infty} \left( T h_c(T) \right) \leq 1
\qquad .
\end{eqnarray}

\subsection{ Thermodynamic quantities }

Following the computations done for the symmetric case 
in Section \ref{thermosym},
we obtain that the energy per monomer $e(T)$ of the chain
in the thermodynamic limit
can be obtained as
\begin{eqnarray} \label{ebiased}
e(T)= -
\frac{\int_{Q,l} Q [P^+_{\Gamma_{eq}(T)}(Q,l)+P^-_{\Gamma_{eq}(T)}(Q,l)]   }{ \int_{Q,l} l [P^+_{\Gamma_{eq}(T)}(Q,l)+P^-_{\Gamma_{eq}(T)}(Q,l)] }
\qquad ,
\end{eqnarray}
with $\Gamma_{eq}(T)$ defined in Equation (\ref{geqd}).

Using the fixed point solution of Equations (\ref{solu-biased}), we have
at large temperature 
\begin{eqnarray}
e(T) && \simeq  - 2 \delta \sigma \coth [\delta \Gamma_{eq}(T)]
- \frac{2 \delta^2  \sigma \Gamma_{eq}(T) }{ \sinh^2 [\delta \Gamma_{eq}(T)]}
\nonumber  \\
&& \simeq - 2 \delta \sigma 
-8 \sigma \delta^2 \Gamma_{eq}(T) e^{-2 \delta \Gamma_{eq}(T)}+ \ldots
\end{eqnarray}
Since in the limit of small dissymmetry $\delta \to 0$,
we have the relation (\ref{univ}), the energy per monomer at the transition $T_c$ is simply 
\begin{eqnarray}
e(T_c)= - q_0  
\qquad ,
\end{eqnarray}
and coincides of course with the energy per monomer 
 when the heteropolymer
is delocalized in the solvent $(+)$ for $T>T_c$.
Surprisingly however, we find that the critical behavior
near $T_c$ is governed by the following essential singularity :
\begin{eqnarray}
e(T)-e(T_c)  && \simeq - 8 \sigma \delta^2 \Gamma_{eq}(T) e^{-2 \delta \Gamma_{eq}(T)}  \nonumber \\
&&  \opsimeq_{T \to T_c^-}  - 2 q_0 
\frac{\ln \frac{4 \sigma}{q_0}}{( 1-\frac{T}{T_c} )} 
\exp \left[ - \frac{\ln \frac{4 \sigma}{q_0}}{( 1-\frac{T}{T_c} )} \right]
\qquad .
\end{eqnarray}

The entropy per monomer $s(T)$
is given in the thermodynamic limit by the generalization of 
Equation (\ref{ssymm})
\begin{eqnarray} 
s(T) =
&& \frac{\int_{Q,l} [l \ln \mu 
+\ln \kappa -\frac{3}{2} \ln l ] 
[P^+_{\Gamma_{eq}(T)}(Q,l)+P^-_{\Gamma_{eq}(T)}(Q,l)]  }{ \int_{Q,l} l 
[P^+_{\Gamma_{eq}(T)}(Q,l)+P^-_{\Gamma_{eq}(T)}(Q,l)]} 
\nonumber \\
&&=  \ln \mu- \frac{3}{2}  \frac{ \overline{\ln l^+} + \overline{\ln l^-}}
{\overline{ l^+} + \overline{ l^-}}
- 2 \frac{\ln \kappa} {\overline{ l^+} + \overline{ l^-}}
\qquad ,
\end{eqnarray}
 where the averages have to be computed with the
probability distributions $P_{\Gamma}^{\pm}(Q,l)$
given in Equations (\ref{solu-biased}). 
Considering as before 
 the regime $\delta \to 0$, $\Gamma \to \infty$, 
with $\gamma=\delta \Gamma \gg 1$ fixed 
where we can use Equation (\ref{asymp+}),
we get 
\begin{eqnarray} 
s(T) && \simeq  \ln \mu - 
12 \sigma  \delta^3  \Gamma_{eq}(T) e^{-2 \delta \Gamma_{eq}(T)} +\ldots 
\nonumber  \\
&&  \opsimeq_{T \to T_c^-}
\ln \mu -\frac{3 q_0^2}{2 \sigma}
\frac{\ln \frac{4 \sigma}{q_0}}{( 1-\frac{T}{T_c} )} 
\exp \left[ - \frac{\ln \frac{4 \sigma}{q_0}}{( 1-\frac{T}{T_c} )} \right]
\qquad .
\end{eqnarray}

We finally give the expression of the free-energy per monomer
\begin{eqnarray} \label{freedyss}
f(T) && \simeq f(T_c) 
- 8 \sigma \delta^2 \Gamma_{eq}(T) e^{-2 \delta \Gamma_{eq}(T)}+
12 \sigma  \delta^3 T \Gamma_{eq}(T) e^{-2 \delta \Gamma_{eq}(T)}... 
\nonumber \\
&& \opsimeq_{T \to T_c^-}  f(T_c)
- 2 q_0 \left(\ln \frac{4 \sigma}{q_0}  \right) 
\exp \left[ - \frac{ \ln \frac{4 \sigma}{q_0} }
{ \left(1-\frac{T}{T_c} \right)} \right]
\end{eqnarray}
which again presents an essential singularity near $T_c$ :
the delocalization transition is thus found to be of infinite order.

The well known {\it pure} models presenting infinite order transitions
are the two-dimensional XY model \cite{KT}
and the 1D Ising or Potts model with inverse square long range interactions 
\cite{anderson} \cite{cardy}. 
The important property of these systems is that
 the interaction between two defects (two vortices
in the XY model or two kinks in the spin chain) is
 logarithmic at larges distances. The usual 
estimation of the free-energy for a pair of defects, where both
the energy and the entropy behave as the logarithm of the size $L$
of the system, leads to a factorization of $\ln L$ as 
in the Imry-Ma argument leading to Equation (\ref{relationflp})
for the heteropolymer problem. It would be interesting to 
discuss in more details the similarities/differences
between these systems, but this goes beyond the scope of the present paper.

\subsection{ Statistical properties of blobs}

In the localized phase $T<T_c$, we find that the blobs in the domain $z>0$
and $z<0$ have respectively the typical lengths (\ref{lengthpm})
\begin{eqnarray} \label{lpmblobs}
&& l^{+}_{ blob}(T)  \simeq \overline{l^+}_{\Gamma_{eq}(T)}
\simeq \frac{1}{4 \delta^2 \sigma } (e^{2\delta \Gamma_{eq}(T)} 
-2  \delta \Gamma_{eq}(T) -1 )  \\
&& l^{-}_{ blob}(T)  \simeq \overline{l^-}_{\Gamma_{eq}(T)}
\simeq  \frac{1}{4 \delta^2 \sigma } ( 2\delta \Gamma_{eq}(T)- 1+
e^{- 2 \delta \Gamma_{eq}(T)}  )
\qquad .
\end{eqnarray}
As $T$ approaches $T_c^-$, $l^{+}_{ blob}(T)$
thus diverges with an
essential singularity as
\begin{eqnarray} \label{lplusT}
l^{+}_{ blob}(T)  \opsimeq_{T \to T_c^-}
 \frac{1}{ 4 \delta^2 \sigma}
\left( \frac{1}{\delta^2 \sigma} \right)^{\frac{ 1  }{  (1-\frac{T}{T_c} )} }
= \frac{\sigma}{q_0^2}  \exp \left[ + \frac{ \ln \frac{4 \sigma}{q_0} }
{ \left( 1-\frac{T}{T_c} \right)} \right] 
\qquad ,
\end{eqnarray}
whereas $l^{-}_{ blob}(T)$ also diverges but only algebraically as
\begin{eqnarray} \label{lmoinsT}
l^{-}_{ blob}(T)  \opsimeq_{T \to T_c^-} 
\frac{\ln \frac{1}{\delta^2 \sigma}}{ 4 \delta^2 \sigma} \frac{1}{(1-\frac{T}{T_c} )}
= \frac{\sigma}{q_0^2} \frac{ \ln \frac{4 \sigma}{q_0} }
{\left(1-\frac{T}{T_c} \right)} 
\qquad .
\end{eqnarray}

We note that these behaviors are compatible with the relation
(\ref{relationlmlp})
between the typical lengths $l^{-}_{blob}$ and $l^{+}_{blob}$
obtained in Reference \cite{garel} via an Imry-Ma argument,
since we have obtained
\begin{eqnarray} \label{relationlblobs}
l^{-}_{ blob}  \simeq 
\frac{\sigma}{q_0^2} \ln l^{+}_{ blob} 
\qquad .
\end{eqnarray}
Our results (\ref{freedyss},\ref{lplusT}) also satisfy
the relation (\ref{relationflp}) between the free-energy
and the typical blob length $l^{+}_{ blob}$ in the preferred
solvent obtained again in Reference \cite{garel} via an Imry-Ma argument,
since we have
\begin{eqnarray} \label{relationflplus}
f(T) -f(T_c) \simeq 
 - \frac{3}{2} (T_c-T) \frac{ \ln l^{+}_{ blob} }{ l^{+}_{ blob}}
\end{eqnarray}

As a direct consequence of Equations (\ref{lplusT}, \ref{lmoinsT}),
 we obtain
that the typical distances $z^{\pm}(T)$
to the interface in the domain $z>0$ and $z<0$
diverge respectively as $T \to T_c^-$ as 
\begin{eqnarray}
&& z^+(T) \sim \sqrt{ D l^{+ blob}(T)} \opsimeq_{T \to T_c^-}
 = \frac{ \sqrt{ D \sigma}}{q_0}  \exp \left[  \frac{ \ln \frac{4 \sigma}{q_0} }
{ 2 \left( 1-\frac{T}{T_c} \right)} \right]
 \\
&& z^-(T) \sim \sqrt{ D l^{- blob}(T)} 
\opsimeq_{T \to T_c^-} 
= \frac{ \sqrt{ D \sigma}}{q_0} \left[ \frac{ \ln \frac{4 \sigma}{q_0} }
{\left(1-\frac{T}{T_c} \right)} \right]^{1/2}
\qquad .
\end{eqnarray}

As in the symmetric case, the lengths $l^{\pm}$ of the blobs
are independent random variables distributed respectively with
the probability distributions $P^{\pm}_{\Gamma_{eq}(T)}(l)$.
In particular, near $T_c$, the rescaled variable 
\begin{eqnarray} 
\lambda_+= 4 \delta^2 \sigma \frac{ l } {  e^{ 2 \delta \Gamma_{eq}(T)}}
\opsimeq_{T \to T_c^-} \frac{l_+}
{ \frac{\sigma}{q_0^2} 
 \exp \left[ \frac{ \ln \frac{4 \sigma}{q_0} }  
{ \left( 1-\frac{T}{T_c} \right)} 
\right] } 
\end{eqnarray}
is distributed with a simple exponential law $e^{-\lambda_+}$
(see Equation \ref{asymp+}).

\subsection{Density profiles $\rho^{\pm}(z)$ }

Following the calculations done previously in the symmetric
case, we obtain that the probabilities $\rho^{\pm}(z)$ 
that a given point of the polymer
is at a distance $z$ from the interface in the $(\pm)$ domain
(with the normalization $\int_0^{\infty} dz (\rho^+(z)+ \rho^-(z))=1$)
reads
\begin{eqnarray}
 \rho^{\pm}(z) = 
   \frac{1}
{\int_0^{\infty} dl \ l \ (P^+_{\Gamma}(l)+P^-_{\Gamma}(l) )}
\int_0^{\infty} dl P^{\pm}_{\Gamma}(l) \ 
4 \sqrt{\frac{l}{ 2 D}} \int_{\frac{2z}{\sqrt{ 2 D l}}}^{\infty} du e^{-u^2}
\qquad .
\end{eqnarray}

In the regime $\delta \Gamma \gg 1$
studied above where $ P^{+}_{\Gamma}(l) $
has the asymptotic simple form of Equation (\ref{asymp+}), we get
the following scaling form
\begin{eqnarray}
\rho^+(z) dz = R^+(Z) dZ 
\qquad ,
\end{eqnarray}
where the rescaled distance $Z$ to the interface 
and the scaling function $R(Z)$ read
\begin{eqnarray}
 Z=  \frac{ 4 \delta z}{ \sqrt{ 2 D} e^{\delta \Gamma_{eq}(T)}}
\opsimeq_{T \to T_c^-}
 \frac{  z}{ \sqrt{ \frac{ D}{2} } \frac{\sigma}{q_0} 
\exp \left[ + \frac{ \ln \frac{4 \sigma}{q_0} }
{ 2 \left( 1-\frac{T}{T_c} \right)} \right] }
\qquad ,
\end{eqnarray}
and 
\begin{eqnarray}
R^+(Z)= 4 \int_0^{\infty} du e^{- u^2}
\int_{\frac{Z}{u}}^{\infty} dv v^2 e^{- v^2}
\qquad .
\end{eqnarray}
Deriving with respect to $Z$ yields
\begin{eqnarray}
\frac{d R^+(Z)}{dZ}= - 4 Z^2  \int_0^{\infty} du e^{- u^2-\frac{Z^2}{u^2}} \frac{1}{u^3}
= -4 Z K_1( 2 Z)
\qquad .
\end{eqnarray}
In particular, the asymptotic behaviors at $Z \to 0$ and $Z \to \infty$
are given respectively by
\begin{eqnarray}
 R^+(Z)= \frac{\pi}{2} -2 Z +O(Z^2) 
\end{eqnarray}
and
\begin{eqnarray}
R^+(Z)\opsimeq_{Z \to \infty}
 \sqrt{\pi Z} e^{-2 Z} 
\qquad .
\end{eqnarray}

\subsection{Correlation function 
$\overline{< {\rm sgn} (z(s)) {\rm sgn} (z(s')) > }$}

Within the renormalization picture, the correlation function 
$\overline{< {\rm sgn} (z(s)) {\rm sgn} (z(s')) > }$ 
of the solvents seen by two monomers at distance $(s-s')$
along the polymer chain containing a dissymmetric distribution
in hydrophilic/hydrophobic components
corresponds exactly to the spin correlation function
$\overline{\langle S_0(t)S_{x}(t) \rangle}$ of 
the random field Ising model in the Glauber
dynamics starting from a random initial condition
in the presence of an external field, which is computed in 
Reference \cite{RGRFIM}.
Here we do not give the full result, but only
the correlation length $\xi(T)$ 
which, not surprisingly, is simply given in the regime
$\delta \Gamma \gg 1$ that we consider by
the typical behavior of $l^{+}_{ blob}(T)$ obtained in Equation 
(\ref{lpmblobs})
\begin{eqnarray} 
\xi(T) \simeq l^{+}_{ blob}(T) \opsimeq_{T \to T_c}
\frac{\sigma}{q_0^2}  \exp \left[ + \frac{ \ln \frac{4 \sigma}{q_0} }
{ \left( 1-\frac{T}{T_c} \right)} \right]\qquad .
\end{eqnarray}

\section{Finite size properties }

\label{finite}

Up to now we have always considered the thermodynamic limit 
of heteropolymers of infinite length $L \to \infty$.
In this section, we study the finite-size properties
of the localization at the interface for the case
of periodic boundary conditions, i.e. 
 we consider cyclic chains $(q_1,q_2, \ldots q_L,q_{L+1}=q_1)$
of finite (but large) size $L$, with positions $(z_1, \ldots , z_{L+1}=z_1)$.
We can apply the renormalization procedure as before, but
the procedure will now stop at some finite renormalization
scale $\Gamma_{\rm deloc.}$ where the chain delocalizes
from the interface. This renormalization
scale depends upon the realization of the quenched charges
and corresponds to some temperature $T_{\rm deloc}$
via the relation 
\begin{eqnarray}
\Gamma_{\rm deloc}=\Gamma_{eq}(T_{\rm deloc})
\end{eqnarray}
given in Equations (\ref{gammaeqt}) and (\ref{geqd}) respectively for the symmetric
case and the biased case.
In the following, we characterize the distribution of $T_{\rm deloc}$
for the ensemble of finite chains of size $L$, and the probability
distribution of the number of blobs for the chains
of length $L$ that are still localized at a given temperature.

\subsection{Probability measure for the blobs of a cyclic chain of length $L$}

As shown in References \cite{fisheryoung},\cite{RGSinai},
it is possible to follow the renormalization procedure 
for finite size systems. Here we deal with periodic
boundary conditions, and to avoid problems
with the translation invariance along the chain, it
is convenient to mark a point of the chain,
called ``the origin" in the following, and 
to give an orientation to the chain.
For $k=1,2 \ldots$, we define
 $N_{\Gamma,L}^{2k,\pm}(Q_1,l_1;Q_2,l_2; \ldots Q_{2k},l_{2k})$
 as the probability that in the
chain of length $L$ at renormalization scale $\Gamma$,
the origin belongs to a blob $(Q_1,l_1)$ in the domain $(\pm)$,
and that there are exactly 
$(2k-1)$ other blobs in the chain of absolute charges and lengths 
given by the sequence $(Q_2,l_2; \ldots Q_{2k},l_{2k})$.
We also need to introduce the probability $N_{\Gamma,L}^{1 ,\pm}(Q)$
that the chain at scale $\Gamma$ is already delocalized in the 
 $(\pm)$ solvent with absolute charge $Q$. The normalization of these probabilities read
\begin{eqnarray}  \label{measure}
&& \sum_{k=1}^{\infty} \int_{Q_i \geq \Gamma,l_i} 
N_{\Gamma,L}^{2k +}(Q_1,l_1;Q_2,l_2; \ldots Q_{2k},l_{2k})
\nonumber \\
&& + \sum_{k=1}^{\infty} \int_{Q_i \geq \Gamma,l_i} 
N_{\Gamma,L}^{2k -}(Q_1,l_1;Q_2,l_2; \ldots Q_{2k},l_{2k})
\nonumber  \\
&& + \int_0^{\infty} d Q N_{\Gamma,L}^{1 +}(Q) +
 \int_0^{\infty} d Q N_{\Gamma,L}^{1 -}(Q)= 1
\qquad .
\end{eqnarray}
Note that in the last two terms, the absolute charge $Q$ is a random
variable of the domain $[0,+\infty[$, contrary to the other terms
where by definition of the renormalization rule we have $Q_i \geq \Gamma$.

The renormalization equations for these probabilities read for $k=1,2, \ldots$
\begin{eqnarray} \label{rgmeasure}
&& \partial_\Gamma   
N_{\Gamma,L}^{2k, \pm}(Q_1,l_1;Q_2,l_2; \ldots Q_{2k},l_{2k}) = 
\nonumber \\
&& \sum_{i=2}^{2k} \int_{Q+Q''-\Gamma=Q_{i},l+l'+l''=l_{i}} 
N_{\Gamma,L}^{2k+2, \pm} (Q_1,l_1; \ldots ;Q_{i-1},
l_{i-1};Q,l;\Gamma,l';Q'',l''; Q_{i+1},l_{i+1 }
\ldots ; Q_{2k},l_{2k})  \nonumber \\
&&  \int_{Q+Q''-\Gamma=Q_1,l+l'+l''=l_1} 
N_{\Gamma,L}^{2k+2, \pm} (Q,l;\Gamma,l';Q'',l''; Q_{2},l_{2 }
\ldots ; Q_{2k},l_{2k})  \nonumber \\
&& +  \int_{Q+Q''-\Gamma=Q_1,l+l'+l''=l_1} 
N_{\Gamma,L}^{2k+2, \pm}
( Q'',l'' ; Q_2,l_2; \ldots;
 Q^-_{2k},l^-_{2k};Q,l; \Gamma,l')  \nonumber
\\
&& +  \int_{Q+Q''-\Gamma=Q_1,l+l'+l''=l_1} 
N_{\Gamma,L}^{2k+2, \mp}
(\Gamma,l'; Q'',l'' ; Q_2,l_2; \ldots;
 Q^-_{2k},l^-_{2k};Q,l)  
\qquad ,
\end{eqnarray}
and 
\begin{eqnarray} \label{eqn1}
 \partial_\Gamma   
N_{\Gamma,L}^{1 ,\pm } (Q)
= \int_{l_1,l_2} N_{\Gamma,L}^{2, \pm} (Q +\Gamma,l_1; \Gamma, l_2)
+ \int_{l_1,l_2} N_{\Gamma,L}^{2, \mp} (\Gamma,l_1; Q+\Gamma, l_2)
\qquad .
\end{eqnarray}

As already obtained in References \cite{fisheryoung} \cite{RGSinai}
for the case of fixed boundary conditions, 
the above renormalization equations concerning
periodic boundary conditions
are solved by a quasi-factorized form for $k=1,2, \ldots$
\begin{eqnarray} \label{solu-measure}    
&& N_{\Gamma,L}^{2k , \pm }(Q_1,l_1;Q_2,l_2; \ldots Q_{2k},l_{2k}) 
\nonumber \\ 
&& = l_1 P_{\Gamma}^{\pm} (Q_1,l_1) P_{\Gamma}^{\mp} (Q_2,l_2) \ldots
 P_{\Gamma}^{\pm} ( Q_{2k-1},l_{2k-1}) P_{\Gamma}^{\mp} ( Q_{2k},l_{2k})  
\delta\left(L-\sum_{i=1}^{2k} l_i\right)
\qquad ,
\end{eqnarray}
where 
$P_{\Gamma}^{\pm}(Q,l)$ are the bulk distributions 
given in Equation (\ref{solu-biased}), satisfying
\begin{eqnarray}   \label{rgP}
 \partial_\Gamma  P_\Gamma^{\pm} (Q,l)  =
&& \int_{Q_1+Q_3-\Gamma=Q,l_1+l_2+l_3=l}
P_\Gamma^{\mp}(\Gamma,l_2) P_\Gamma^{\pm}(Q_1,l_1) 
 P_\Gamma^{\pm}(Q_3,l_3)
\nonumber \\
 && + P_\Gamma^{\pm} (Q,l) \int_0^{\infty} dl' \left(P_\Gamma^{\pm}(\Gamma,l') 
- P_\Gamma^{\mp}(\Gamma,l')\right)
\qquad .
\end{eqnarray}
Note that in the measure (\ref{solu-measure}), the first bond
plays a special role as it is defined as the bond containing the origin.

The equation (\ref{eqn1}) now reads
\begin{eqnarray} \label{eqn1ex}
 \partial_\Gamma   N_{\Gamma,L}^{1 ,\pm } (Q)
=&&  \int_{l_1,l_2} l_1 P_\Gamma^{\pm}(Q +\Gamma,l_1) P_\Gamma^{\mp}( \Gamma, l_2)
\delta\left(L-(l_1+l_2) \right) 
\nonumber \\
&& + \int_{l_1,l_2} l_1 P_\Gamma^{\mp}( \Gamma,l_1 ) P_\Gamma^{\pm} ( Q+\Gamma, l_2) \delta\left(L-(l_1+l_2) \right) 
\nonumber \\
 = && L   P_\Gamma^{\pm}(Q +\Gamma,.) *_L P_\Gamma^{\mp}( \Gamma, .) 
\qquad .
\end{eqnarray}
Using now the explicit expressions (\ref{solu-biased}), we get
in Laplace transform with respect to $L$
\begin{eqnarray} 
\int_0^{\infty} dL e^{-pL} 
\left( \partial_\Gamma   N_{\Gamma,L}^{1 , \pm } (Q) \right) =
  -\partial_p 
\left( \frac{\Delta^2(p)}{\sinh^2[\Gamma \Delta(p)]} 
e^{-Q (\Delta(p) \coth{[\Gamma \Delta(p)]} 
\mp \delta )} \right)  
\end{eqnarray}
where $\Delta(p)  =  \sqrt{\delta^2+\frac{p}{\sigma}}$.

\subsection{Probability for a chain to be delocalized at renormalization scale $\Gamma$}

The equations (\ref{eqn1ex}) for the probabilities
$N_{\Gamma,L}^{1 , \pm } (Q)$ that a chain of length $L$
is already delocalized at scale $\Gamma$ with an absolute
charge $Q$ in the domain $(\pm)$ can be integrated with respect to $\Gamma$
to yield
\begin{eqnarray} 
\int_0^{\infty} dL e^{-pL} 
   N_{\Gamma,L}^{1 ,\pm } (Q)  =
\frac{1}{2 \sigma} \left(
\frac
{\coth [\Gamma \Delta(p)]}{\Delta(p)}-\frac{\Gamma}{\sinh^2[\Gamma \Delta(p)]} \right) 
e^{-Q (\Delta(p) \coth[\Gamma \Delta(p)] \mp \delta )}
\qquad .
\end{eqnarray}
We may check that $N_{\Gamma ,L}^{1 ,\pm } (Q)$
vanishes as it should for $\Gamma \to 0$.
In the limit $\Gamma \to \infty$, the chain is expected
to be always delocalized, with 
a charge corresponding to the sum $\sum_{i=1}^{L} q_i$
of L independent identical random variables.
Indeed, using (\ref{univ}), we find
\begin{eqnarray} 
  N_{\Gamma \to \infty,L}^{1 , \pm } (Q)
= {\cal L}^{-1}_{p \to L} \left[ 
\frac{ e^{-Q (\sqrt{\delta^2+\frac{p}{\sigma}} 
\mp \delta ) }}{ 2 s \sqrt{\delta^2+\frac{p}{\sigma}}} \right]
= \frac{ 1 }{ 2 \sqrt{2 \sigma L}}
e^{- \frac{(Q \mp L q_0)^2}{4 \sigma L}}
\qquad ,
\end{eqnarray}
as expected from the Central Limit Theorem.

For arbitrary $\Gamma$, we can obtain the total probabilities
$N_{\Gamma ,L}^{1 , \pm }$ that the chain is already delocalized
at scale $\Gamma$ in the domain $(\pm)$ as
\begin{eqnarray} 
&& N_{\Gamma,L}^{1 , \pm } = \int_0^{\infty} dQ
   N_{\Gamma,L}^{1 , \pm } (Q) 
\nonumber \\
&& =
{\cal L}^{-1}_{p \to L} \left[ 
\frac{1}{2 \sigma (\Delta(p) \cosh{[\Gamma \Delta(p)]} 
\mp \delta \sinh{[\Gamma \Delta(p)]} ) } \left(
\frac{\cosh[\Gamma \Delta(p)]}{\Delta(p)}
-\frac{\Gamma}{\sinh[\Gamma \Delta(p)]} \right)
\right]
\qquad .
\end{eqnarray}

In particular for the symmetric case $\delta=0$, we have
\begin{eqnarray}  
 N_{\Gamma,L}^{1 ,\pm } = \frac{1}{2} {\cal N} 
\left( \lambda \equiv \frac{ \sigma L}{ \Gamma^2}\right) 
\qquad .
\end{eqnarray}
The scaling function ${\cal N} ( \lambda )$ represents
the total probability for a symmetric chain of length $L$
to be delocalized at temperature $T$ corresponding
to the renormalization scale $\Gamma=\Gamma_{eq}(T)$ 
defined in Equation (\ref{gammaeqt}).
It reads
\begin{eqnarray}  
 {\cal N} ( \lambda ) && ={\cal L}^{-1}_{s \to \lambda} 
\left( \frac{1}{s} - \frac{2}{ \sqrt s \sinh (2 \sqrt s ) }\right)  
\nonumber \\
&&= 1- \sum_{n=-\infty}^{+\infty} (-1)^n e^{- \lambda n^2 \frac{\pi^2}{4}}
= 1- \frac{2}{\sqrt{\pi \lambda}} 
\sum_{m=-\infty}^{+\infty}  e^{-  \frac{(2m+1)^2}{\lambda}}
\qquad .
\end{eqnarray}

\subsubsection{Distribution of the delocalization temperature 
for symmetric finite chains}

The probability that the polymer chain delocalizes from the interface
between $\Gamma$ and $\Gamma+d\Gamma$ is given by 
$(\partial_{\Gamma} [ {\cal N} ( \lambda ) ] )$. 
To characterize the distribution of
 the delocalization temperature $T_{\rm deloc}$
for the ensemble of the symmetric finite chains of length
 $L$, it is thus convenient
to define the rescaled variable
\begin{eqnarray}  
g=\frac{\Gamma_{eq}(T_{\rm deloc})}{\sqrt{\sigma L}}
\qquad ,
\end{eqnarray}
where the function $\Gamma_{eq}(T)$ has been defined in Equation (\ref{gammaeqt}).
The final result is that $g$ is distributed with the law :
\begin{eqnarray}  
D(g)  = - \frac{2}{g^3} {\cal N}' \left( \frac{1}{g^2} \right) 
&& = \frac{\pi^2}{g^3}\sum_{n=1}^{+\infty} (-1)^{n+1} n^2 e^{-   \frac{n^2 \pi^2}{4 g^2}} \\
&& = \frac{2}{\sqrt{\pi }} 
\sum_{m=-\infty}^{+\infty} 
\left[ 2 (2m+1)^2 g^2 -1 \right] e^{-  (2m+1)^2 g^2}
\qquad .
\end{eqnarray}

\subsubsection{Distribution of the delocalization temperature 
for dissymmetric finite chains}

In the biased case, the probability that a chain of length $L$
delocalizes from the interface between $\Gamma$ and $\Gamma+d\Gamma$
in the domain $(\pm)$ is given by 
\begin{eqnarray} 
 \partial_\Gamma   N_{\Gamma,L}^{1 ,\pm }  =
 {\cal L}^{-1}_{p \to L} \left[  -\partial_p 
\left( \frac{\Delta^2(p)}{\sinh[\Gamma \Delta(p)]
(\Delta(p) \cosh{[\Gamma \Delta(p)]} \mp \delta \sinh[\Gamma \Delta(p)])} \right)   \right] 
\qquad .
\end{eqnarray}
In the regime $\gamma=\delta \Gamma \gg 1$ considered before 
(see Equation (\ref{asymp+})), the inverse Laplace transform is dominated
by a single pole contribution
\begin{eqnarray} 
 \partial_\Gamma   N_{\Gamma,L}^{1 , +}  \simeq
2 \delta a^+(\gamma) L e^{- L a^+(\gamma) }
\qquad ,
\end{eqnarray}
where $a^+(\gamma)= 4 \sigma \delta^2 \exp(-2 \gamma)$ has been introduced in (\ref{asymp+}). We now use the correspondence 
between temperature and
renormalization scale given by $\Gamma=\Gamma_{eq}(T)$ (\ref{geqd}).
To characterize the distribution of
 the delocalization temperature $T_{\rm deloc}$
for the ensemble of dissymmetric finite chains of large length
 $L$, it is thus convenient
to define the random variable
\begin{eqnarray}  
r=\frac{1}{4 \delta^2 \sigma L} \left( \frac{1}{\delta^2 \sigma} \right)^{\frac{T_{\rm deloc}}{T_c-T_{\rm deloc}}}
\qquad .
\end{eqnarray}
The final result is that $r$ is distributed with the law
\begin{eqnarray}  
D^+(r)= \frac{1}{r^2} e^{-\frac{1}{r}} 
\qquad .
\end{eqnarray}
In particular, the typical value for the delocalization temperature
corresponds to $r \sim 1$ and thus presents a correction in $1/(\ln L)$
with respect to the critical temperature $T_c$ 
\begin{eqnarray}  
 T_{\rm deloc}^{typ} \sim T_c \left(1-\frac{\frac{1}{\delta^2 \sigma}}
{ \ln L} \right)
\qquad .
\end{eqnarray}

\subsection{Distribution of the number of blobs at scale $\Gamma$}

The probability that a chain is still localized around the interface 
at scale $\Gamma$ and is in a state with $(2k)$ blobs read
for $k=1,2, \ldots$ 
\begin{eqnarray}  
B^{2k}_L(\Gamma) =  \int_{Q_i,l_i} 
N_{\Gamma,L}^{2k , +}(Q_1,l_1;Q_2,l_2; \ldots Q_{2k},l_{2k}) 
\nonumber \\
+  \int_{Q_i,l_i} 
N_{\Gamma,L}^{2k , -}(Q_1,l_1;Q_2,l_2; \ldots Q_{2k},l_{2k}) 
\qquad .
\end{eqnarray}
The Laplace transform of the generating function 
of the probabilities $ B^{2k}_L(\Gamma)$ reads
\begin{eqnarray}  
\int_0^{\infty} dL e^{-pL} 
\left( \sum_{k=1}^{\infty} z^k B^{2k}_L(\Gamma)\right) 
&& = \left[ -\partial_p \left( P^+_{\Gamma}(p) P^-_{\Gamma}(p) \right) \right]
 \frac{z }{1- z P^+_{\Gamma}(p) P^-_{\Gamma}(p)} 
\nonumber \\
&& =  \frac{z \sinh [\Gamma \Delta(p)] (\delta^2 \sigma \sinh [\Gamma \Delta(p)]
 +p \Gamma \Delta(p) \cosh [\Gamma \Delta(p)] )}
{ (\delta^2 \sigma (1-z) +p  (\cosh^2 [\Gamma \Delta(p)]-z) ) 
(\delta^2 \sigma +p  \cosh^2 [\Gamma \Delta(p)])}
\ .
\end{eqnarray}

In particular, the average number of blobs 
reads
\begin{eqnarray} 
<2k && >_{L,\Gamma}  \equiv  \sum_{k=1}^{\infty} 2k B^{2k}_L(\Gamma) 
 =  {\cal L}^{-1}_{p \to  L} \left(
2 \left[ -\partial_p \left( P^+_{\Gamma}(p) P^-_{\Gamma}(p) \right) \right]
 \frac{1 }{(1-  P^+_{\Gamma}(p) P^-_{\Gamma}(p))^2} \right)  
\nonumber \\
 = && {\cal L}^{-1}_{p \to  L} \left(
2 \frac{\delta^2 \sigma +p \Gamma \Delta(p) \coth [\Gamma \Delta(p)]}
{p^2  \sinh^2 [\Gamma \Delta(p)]} \right) 
\nonumber \\
= && \frac{2 \sigma L \delta^2}{ \sinh^2(\delta \Gamma)}
+ \sum_{n=1}^{\infty} e^{- \sigma L (\delta^2+\frac{n^2 \pi^2}{\Gamma^2})}
4 \frac{\sigma L n^2 \pi^2}{\Gamma^2 (n^2 \pi^2+\delta^2 \Gamma^2)} 
\left( 1+\frac{2 \delta^2 \Gamma^2}{n^2 \pi^2+\delta^2 \Gamma^2}-2 \frac{\sigma L n^2 \pi^2}{\Gamma^2} \right)
\ .
\end{eqnarray}
The first term is proportional to $L$ and corresponds of course
to the ratio $\frac{2 L}{\overline{l_{\Gamma}^+}+\overline{l_{\Gamma}^-}}$
that could have been anticipated from the study of Section
\ref{biased} concerning the thermodynamic
limit $L \to \infty$. 
 The other terms, which represent the finite-size corrections 
to this dominant contribution, decay exponentially with $L$.

We may also compute the second moment
\begin{eqnarray} 
&& <(2k)^2>_{L,\Gamma}  \equiv  \sum_{k=0}^{\infty} 4 k^2 B^{2k}_L(\Gamma) 
=  
{\cal L}^{-1}_{p \to   L} \left(
4 \left[ -\partial_p \left( P^+_{\Gamma}(p) P^-_{\Gamma}(p) \right) \right]
 \frac{1+P^+_{\Gamma}(p) P^-_{\Gamma}(p) }{(1-  P^+_{\Gamma}(p) P^-_{\Gamma}(p))^3} \right)  \nonumber \\
&& =  {\cal L}^{-1}_{p \to  L} \left(
4 \frac{(2 \delta^2 \sigma +p (\cosh^2 [\Gamma \Delta(p)]+1))
(\delta^2 \sigma +p \Gamma \Delta(p) \coth [\Gamma \Delta(p)])}
{p^3  \sinh^4 [\Gamma \Delta(p)]} \right) 
\nonumber \\
&& = \frac{4 \sigma^2 L^2 \delta^4}{ \sinh^4(\delta \Gamma)}
+\frac{2 \sigma L \delta^2}{ \sinh^4(\delta \Gamma)}
\left( \cosh(2 \delta \Gamma) - 4 \delta \Gamma \coth( \delta \Gamma)+3 \right)
+ \sum_{n=1}^{\infty} e^{- \sigma L (\delta^2+\frac{n^2 \pi^2}{\Gamma^2})}
( \ldots )
\qquad .
\end{eqnarray}
The dominant behavior of the variance of the number of blobs
is thus given by 
\begin{eqnarray} 
<(2k)^2>_{L,\Gamma} - (<2k>_{L,\Gamma})^2 =
\frac{2 \sigma L \delta^2}{ \sinh^4(\delta \Gamma)}
\left( \cosh(2 \delta \Gamma) - 4 \delta \Gamma \coth( \delta \Gamma)+3 \right)
+ ...
\end{eqnarray}

In the symmetric case $\delta=0$, the generating function 
may be more explicitly computed as
\begin{eqnarray} 
 \sum_{k=1}^{\infty} z^k B^{2k}_L(\Gamma)  
= &&  {\cal L}^{-1}_{p \to L} \left(
\frac{ \Gamma z {\rm \tanh } [\Gamma \Delta(p)]}
{ \sqrt{ p \sigma} ( \cosh^2 [\Gamma \Delta(p)] -z)} \right)  
\nonumber \\
&& =\sum_{n=-\infty}^{+\infty} \left(
e^{-\frac{\sigma L}{\Gamma^2} (\alpha+n\pi)^2}
- e^{-\frac{\sigma L}{\Gamma^2} (\frac{\pi}{2}+n\pi)^2} \right)
\qquad ,
\end{eqnarray}
where $\alpha ={\rm ArcCos} \sqrt{z} \in (0,\frac{\pi}{2})$ for $z \in (0,1)$.
In particular, the average and the variance of the number of blobs read :
\begin{eqnarray} 
&& <2k>_{L,\Gamma}   = 2 \frac{\sigma L}{\Gamma^2}
+ \sum_{n=1}^{+\infty} 4 \frac{\sigma L}{\Gamma^2}
 \left(1-2 \frac{\sigma L}{\Gamma^2} n^2 \pi^2 \right)
e^{-\frac{\sigma L}{\Gamma^2} n^2 \pi^2}  \\
&& <(2k)^2>_{L,\Gamma} - (<2k>_{L,\Gamma})^2 =
\frac{4 \sigma L }{ 3 \Gamma^2}
+ ...  \qquad ,
\end{eqnarray}
where $\Gamma=\Gamma_{eq}(T)$, and where the dots represent
terms that are exponentially small in $(L/\Gamma^2)$. 

\section{Non-equilibrium dynamics starting from a zero-temperature initial
condition}

\label{dynamics}

As explained in Section \ref{dynamicsinter},
the dynamics at temperature $T$ for $t>0$
starting from a zero-temperature initial condition at $t=0$,
i.e. after a quench to $T=0$ for $t<0$, can be
described by the renormalization procedure, where
the renormalization scale $\Gamma$ now corresponds to time via 
\begin{eqnarray} 
\Gamma(t)=T \ln \frac{t}{t_0} 
\end{eqnarray}
as in References \cite{RGletter} \cite{RGSinai}.
The dynamics takes place 
up to time $t_{eq}$ where equilibrium at temperature $T$ is reached :
\begin{eqnarray} 
T \ln \frac{t_{eq}}{t_0} = \Gamma_{eq}(T) 
\qquad .
\end{eqnarray}
In the symmetric case, we thus have the scaling
\begin{eqnarray} 
t_{eq} \sim \Gamma^3_{eq}(T) \sim T^3 (\ln T)^3
\qquad ,
\end{eqnarray}
whereas in the biased case
\begin{eqnarray} 
t_{eq} \sim \exp \left( \frac{3 \ln \frac{1}{\delta^2 \sigma}}
{ 4 \left(1-\frac{T}{T_c} \right)} \right)
\qquad .
\end{eqnarray}

All the quantities computed before 
for the equilibrium at temperature $T$ as functions
of $\Gamma_{eq}(T)$ have the same expressions
for the dynamics as functions of $\Gamma(t)=T \ln t$
for large time $t<t_{eq}$.
For instance, in the symmetric case, the typical length
of blobs behaves as
\begin{eqnarray} 
l(t) \sim \frac{\Gamma^2(t)}{\sigma} \sim \frac{(T \ln t)^2}{\sigma}
\qquad ,
\end{eqnarray}
whereas in the biased case, the typical lengths
of blobs in the domains $z>0$ and $z<0$ behave
respectively as
\begin{eqnarray} 
&& l^+(t) \sim \frac{1}{ 4 \delta^2 \sigma} e^{2 \delta \Gamma(t)}
\sim \frac{1}{ 4 \delta^2 \sigma} t^{2 \delta T}
= \frac{\sigma}{q_0^2} t^{\frac{q_0 T}{\sigma}}  
\qquad ,\\
&& l^-(t) \sim \frac{ \Gamma(t)}{ 2 \delta \sigma}
\sim \frac{ T \ln t}{ 2 \delta \sigma} = \frac{ T \ln t}{ q_0}
\qquad .
\end{eqnarray}

\section{Conclusion}

\label{conclusion}

In this paper, we have proposed a new approach based on a disorder-dependent
renormalization procedure to study the localization of
random heteropolymers at the interface between two selective solvents
within the model of Garel et al. \cite{garel}.
The renormalization procedure has been defined to 
 construct an effective thermodynamics, where one only considers
the heteropolymer configurations 
that correspond to the optimal Imry-Ma domain structure.
At high temperatures,
where the distribution of absolute charges of blobs 
generated by the renormalization procedure becomes infinitely broad,
the effective thermodynamics is expected to become accurate
and to give asymptotic exact results.
With the renormalization approach,
we have recovered that a chain with a symmetric distribution
in hydrophilic/hydrophobic components is 
localized at the interface at any finite temperature
in the thermodynamic limit, whereas a dissymmetry
in hydrophilic/hydrophobic components leads
to a delocalization phase transition, in agreement
with previous studies. In addition, for both cases, 
we have given explicit expressions 
for the high temperature behaviors
of various physical quantities characterizing the localized phase, 
including in particular the free-energy per monomer,
the distribution of the blob lengths in each solvent and
the polymer density in the direction perpendicular to the interface.
For the case of a small dissymmetry in hydrophobic/hydrophilic components, 
where the delocalization transition takes place at high temperature,
the renormalization approach allows to study
the critical behaviors near the transition :
we have found that
the free energy presents an essential singularity 
at the transition, making all of its derivatives continuous at $T_c$
(infinite order transition),
 that the typical length of blobs in the preferred
solvent diverges with an essential singularity, whereas
the typical length of blobs in the other
solvent diverges algebraically.
We have then studied the finite-size properties
of the problem by considering
cyclic finite (large) chains. In particular, 
we have given the probability distribution
of the delocalization temperature for the ensemble
of chains of fixed finite length $L$, and the distribution
of the numbers of blobs in the chain still localized at
some temperature. 
Finally, we have briefly discussed
 the non-equilibrium dynamics at temperature $T$ 
starting from a zero-temperature initial condition.

In conclusion, the disorder-dependent renormalization approach
yields explicit predictions for the high temperature behaviors
of various physical quantities, that would certainly be
interesting to test precisely by numerical studies.
In particular, the predictions concerning the critical behaviors
near the delocalization transition
are certainly surprising,
and disagree with the Monte Carlo studies
of Reference \cite{sommer} where power laws were found, and with the Replica
Gaussian variational approach of Reference \cite{maritanreplica}
where the transition was found to be of second order. 
 It would thus be particularly worthy
to test numerically more precisely
the order of the transition. We hope that the
predictions of the renormalization approach
concerning the finite-size properties
of the problem presented in Section \ref{finite}
will make easier the comparison with numerical studies.
Finally, we would like to stress that the renormalization
approach yields not only predictions for averages
over the quenched disorder, but also precise
predictions sample by sample. For instance, for a given realization
of the quenched charges, one may compare the typical configurations
of the heteropolymer at equilibrium at temperature $T$
with the blobs structure
obtained by the numerical implementation of the renormalization procedure
up to scale $\Gamma_{eq}(T)$. For the case of cyclic chains, one may
also test numerically sample by sample the renormalization prediction for
the delocalization temperature.

\section{Acknowledgments}

It is a pleasure to thank T. Garel, H. Orland, and E. Guitter
for fruitful discussions, and T. Garel for his 
useful remarks on the manuscript.

\appendix

\section{ Imry-Ma argument for the probability distribution
of blob lengths}

\label{appendixImryMa}

In this Appendix, we recall the determination 
via a direct Imry-Ma argument \cite{garelImryMa}
of the asymptotic behavior
of the probability distribution
of blob lengths in the limit of small lengths, 
and compare it with the renormalization result.

As noted by Garel and Orland \cite{garelImryMa},
one may extend the usual Imry-Ma argument 
for the heteropolymer exposed in Reference \cite{garel}
to obtain not only the scaling of the typical length
of Imry-Ma domains, but also some information
on the probability distribution of the length of domains. 
The idea is as follows \cite{garelImryMa}. 
On one hand, the energy associated to a blob of length $l$
is the sum of $l$ independent random variables
and is thus typically given by 
\begin{eqnarray} 
E_l= - u \sqrt{ 2 \sigma l} 
\qquad ,
\end{eqnarray}
where $u$ is a random variable of order 1, 
that has at least to be positive since
the system tries to use the favorable fluctuations of the disorder
to lower its energy. 
According to \cite{garelImryMa}, the probability distribution
for large $u$ is simply given by the Gaussian of the Central Limit Theorem :
\begin{eqnarray} \label{ularge}
G(u) \opsim_{u \to \infty} e^{-u^2/2}
\qquad .
\end{eqnarray}
The behavior of the distribution $G(u)$ for small $u$
seems however not simple to determine a priori, because 
the small values of $u$ are certainly suppressed
with respect to the Gaussian distribution.
Indeed, if an Imry-Ma domain corresponds to a small value of $u$,
it is very likely that the heteropolymer
will prefer an organization into other Imry-Ma domains.
It seems however difficult
to write precisely the consequences of this effect on the distribution
$G(u)$. On the other hand, the entropy associated
to a blob of length $l$ behaves as $(- \frac{3}{2} \ln l)$
(in \cite{garel}, only powers of $l$ were considered
for the symmetric case
and thus $\ln l$ was replaced by $l^0 \sim 1$, but here for 
a better comparison with the renormalization approach, it is more convenient 
to keep this $\ln l$ dependence).
The minimization of the free energy per monomer
\begin{eqnarray} \label{freeenergyIM}
f(l) \sim - \sqrt{ \frac{ 2 \sigma }{ l }} u + \frac{3}{2} T \frac{\ln l}{l}
\end{eqnarray}
with respect to $l$ leads to
\begin{eqnarray} 
u \sim \frac{3 T } {\sqrt{2 \sigma} } \frac{\ln l}{\sqrt{l}}
\qquad .
\end{eqnarray}
The typical blob length $l^{typ}$ corresponding to $u_{typ} \sim 1$
thus satisfies
\begin{eqnarray} \label{ltypical}
\frac{\sqrt{l^{typ}}} {\ln l^{typ}} \sim \frac{3 T } {\sqrt{2 \sigma} u_{typ}}
\qquad . \end{eqnarray}
Since within the RG approach $l^{typ} = \frac{\Gamma_{eq}^2(T)}{ \sigma}$,
the above equation corresponds to equation (\ref{gammaeqt})
with the choice $u_{typ} \sim \sqrt{2}$
i.e.  a numerical factor of order 1.
Now introducing the rescaled length of Imry-Ma domains
\begin{eqnarray} 
\lambda=\frac{l}{l^{typ}}
\qquad ,
\end{eqnarray}
we find that it is simply related to the random variable $u$ through
\begin{eqnarray} \label{ulambda}
u \sim \frac{1}{\sqrt{  \lambda}}
\qquad .
\end{eqnarray}
The behavior (\ref{ularge}) of the distribution $G(u)$
at large $u$ thus corresponds to the following behavior
of the distribution $P(\lambda)$ at small $\lambda$ :
\begin{eqnarray} 
P(\lambda) \opsim_{\lambda \to 0} \frac{1}{\lambda^{3/2}}
e^{-\frac{C}{\lambda}}
\end{eqnarray}
where $C$ is a constant of order 1. This asymptotic behavior corresponds
exactly to the behavior (\ref{pdelzero}) found via the renormalization
approach.
Concerning the behavior of $G(u)$ for small $u$, 
we explained above why it is difficult to determine
it a priori. Using the relation (\ref{ulambda}),
the asymptotic behavior (\ref{pdelinfty}) found
via the RG approach
\begin{eqnarray} 
P_{RG}(\lambda) \opsim_{\lambda \to \infty}
e^{-\frac{\pi^2}{4}\lambda}
\end{eqnarray}
would correspond to 
\begin{eqnarray} 
G(u) \opsim_{u \to 0} \frac{1}{u^3}
e^{-\frac{C'}{u^2}}
\end{eqnarray}
i.e. a very strong suppression of small values of $u$
for the energies of Imry-Ma domains.

\newpage



\begin{references}

\bibitem{review1} T. Garel, H. Orland and E. Pitard, ``Protein folding and
heteropolymers'', in {\em Spin Glasses
and Random Fields}, A.P. Young (ed.), World Scientific, Singapore
(1997) p. 387-443 ; T.E. Creighton (ed.), {\it Protein Folding}, W.H. Freeman, New York (1992).


\bibitem{garel} 
 T. Garel, D.A. Huse, L. Leibler and H. Orland,
Europhys. Lett. {\bf 8} 9 (1989).


\bibitem{experiments} 
H.R. Brown, V.R. Deline and P.F. Green,
Nature {\bf 341} , 221 (1989);
C.A. Dai, B.J. Dair, K.H. Dai, C.K. Ober, E.J. Kramer, C.Y. Hui and
L.W. Jelinsky, Phys. Rev. Lett. {\bf 73} , 2472 (1994).

 \bibitem{yeung} 
 C. Yeung, A.C. Balazs and D. Jasnow,
Macromolecules, {\bf 25} , 1357 (1992).

\bibitem{sommer}
J.U. Sommer, G. Peng and A. Blumen,
Phys. Rev. E {\bf 53} 5509 (1996);
J. Phys. II France {\bf 6} 1061 (1996);
J. Chem. Phys. {\bf 105} 8376 (1996). 

\bibitem{stepanow}
S. Stepanow, J.U. Sommer, and I.Y. Erukhimovich,
Phys. Rev. Lett. {\bf 81} 4412 (1998).

\bibitem{maritanreplica}
A. Trovato and A. Maritan,
cond-mat/9812321.

\bibitem{maritanbounds}
 A. Maritan, M.P. Riva and A. Trovato,
cond-mat/9901292.

 \bibitem{sinai}
Y.G. Sinai, Theor. Prob. Appl. {\bf 38} , 382 (1993).

 \bibitem{albeverio}
S. Albeverio and X.Y. Zhou, 
J. Stat. Phys. {\bf 53} , 573 (1996).

\bibitem{bolthausen}
E. Bolthausen and F. den Hollander, 
Ann. Prob. {\bf 25} , 1334 (1997).

\bibitem{biskup}
M. Biskup and F. den Hollander, 
Ann. Appl. Prob., to appear.

\bibitem{Ma} 
S. K. Ma, C. Dasgupta and C. K. Hu, 
Phys. Rev. Lett. {\bf 43} 1434 (1979), 
C. Dasgupta and S.K. Ma, 
Phys. Rev. B{\bf 22}, 1305 (1980).

\bibitem{fisher} 
D. S. Fisher, 
Phys. Rev. Lett. {\bf} 69 , 534 (1992)
Phys. Rev. B{\bf 50}, 3799 (1994);
Phys. Rev. B{\bf 51}, 6411 (1995).

\bibitem{fisheryoung}
D.S. Fisher and A.P. Young, Preprint cond-mat/9802246.



\bibitem{hyman}
R.A. Hyman and K. Yang, Phys. Rev. Lett. {\bf 78} 1783 (1997);
R.A. Hyman, K. Yang, R.N. Bhatt, and S.M. Girvin,
Phys. Rev. Lett. {\bf 76} 839 (1997);
R.A. Hyman, Ph. D. Thesis, Indiana University (1997).

\bibitem{spin1}
C. Monthus, O. Golinelli and T. Jolicoeur, Phys. Rev. Lett {\bf 79} 3254
  (1997); Phys. Rev. B {\bf 58} , 305 (1998).


\bibitem{RGletter} 
D.S. Fisher, P. Le Doussal, C. Monthus,
Phys. Rev. Lett. {\bf 80} 3539 (1998).

\bibitem{RGSinai} 
 D.S. Fisher, P. Le Doussal, C. Monthus, cond-mat/9811300, to appear in Phys. Rev. E (1999).

\bibitem{RGreadiff} 
 D.S. Fisher, P. Le Doussal, C. Monthus, cond-mat/9901306, to appear in Phys. Rev. E (1999).


\bibitem{RGRFIM} 
 D.S. Fisher, P. Le Doussal, C. Monthus, in preparation.



\bibitem{phi4}
A.J. Bray, B. Derrida and C. Godreche,
 Europhys. Lett. {\bf 27} 175 (1994);
 A.D. Rutenberg and A.J. Bray Phys. Rev. E {\bf 50} 1900 (1994);
 A.J. Bray and B. Derrida Phys. Rev. E {\bf 51} R1633 (1995);
 B. Derrida, C. Godreche and I. Yekutieli Europhys. Lett. {\bf 12} 385 (1990) and Phys. Rev. A {\bf 44} 6241 (1991).


\bibitem{testRGnume}
A.P. Young and H. Rieger, Phys. Rev. B {\bf 53} , 8486 (1996) ;
A.P. Young, Phys. Rev. B {\bf 56} , 11691 (1997);
F. Igloi and H. Rieger,  Phys. Rev. B {\bf 57} , 11404 (1998) ;
F. Igloi, R. Juhasz and H. Rieger, cond-mat/9811369.

\bibitem{chave}
J. Chave and E. Guitter, J. Phys. A : Math. Gen. {\bf 32} , 445 (1999).

\bibitem{garelImryMa}
T. Garel and H. Orland, private communication.
See T. Garel, G. Iori and H. Orland, Phys. Rev. B {\bf 53} , R2941 (1996)
for the application of the same idea in the context of the random field
XY model. 

\bibitem{ImryMa}
Y. Imry and S.K. Ma, Phys. Rev. Lett. {\bf 35} , 1399 (1975).

\bibitem{KT}
J.M. Kosterlitz and D.J. Thouless, J. Phys. C : Solid State Phys.
{\bf 6} , 1181 (1973) ; J.M. Kosterlitz, J. Phys. C : Solid State Phys.
{\bf 7} , 1046 (1974).

 
\bibitem{anderson} 
P.W. Anderson and G. Yuval, J. Phys. C : Solid State Phys.
{\bf 4} , 607 (1971).

\bibitem{cardy} 
J.L. Cardy, J. Phys. A : Math. Gen. {\bf 14} 1407 (1981).

\end{references}
\end{document}